\documentclass[acmsmall,screen,authorversion,nonacm]{acmart}

\usepackage{changepage}
\usepackage{graphicx}
\usepackage{wrapfig}
\usepackage{multirow}
\usepackage{longtable}

\AtBeginDocument{%
  \providecommand\BibTeX{{%
    \normalfont B\kern-0.5em{\scshape i\kern-0.25em b}\kern-0.8em\TeX}}}

\setcopyright{acmlicensed}
\copyrightyear{2024}
\acmYear{2024}
\acmDOI{XXXXXXX.XXXXXXX}

\acmConference[CHI '25]{ACM CHI Conference on Human Factors in Computing Systems}{April 26--May 01,
  2025}{Yokohama, Japan}
\acmISBN{978-1-4503-XXXX-X/18/06}

\begin{document}

\title[PromptHive: Bringing Subject Matter Experts Back to the Forefront]{PromptHive: Bringing Subject Matter Experts Back to the Forefront with Collaborative Prompt Engineering for Educational Content Creation}

\author{Mohi Reza}
\orcid{0000-0001-9668-3384}
\affiliation{%
  \institution{University of Toronto}
  \city{Toronto}
  \country{Canada}}
\email{mohireza@cs.toronto.edu}
\author{Ioannis Anastasopoulos}
\orcid{0000-0002-1341-5876}
\affiliation{%
  \institution{UC Berkeley}
  \city{Berkeley}
  \country{USA}}
\email{ioannisa@berkeley.edu}
\author{Shreya Bhandari}
\orcid{0009-0007-1705-3052}
\affiliation{%
  \institution{UC Berkeley}
  \city{Berkeley}
  \country{USA}}
\email{shreya.bhandari@berkeley.edu}
\author{Zachary A Pardos}
\orcid{0000-0002-6016-7051}
\affiliation{%
  \institution{UC Berkeley}
  \city{Berkeley}
  \country{USA}}
\email{pardos@berkeley.edu}

\renewcommand{\shortauthors}{Reza et al.}

\begin{abstract}
  Involving subject matter experts in prompt engineering can guide LLM outputs toward more helpful, accurate, and tailored content that meets the diverse needs of different domains. However, iterating towards effective prompts can be challenging without adequate interface support for systematic experimentation within specific task contexts. In this work, we introduce PromptHive, a collaborative interface for prompt authoring, designed to better connect domain knowledge with prompt engineering through features that encourage rapid iteration on prompt variations. We conducted an evaluation study with ten subject matter experts in math and validated our design through two collaborative prompt-writing sessions and a learning gain study with 358 learners. Our results elucidate the prompt iteration process and validate the tool's usability, enabling non-AI experts to craft prompts that generate content comparable to human-authored materials while reducing perceived cognitive load by half and shortening the authoring process from several months to just a few hours.
\end{abstract}


\begin{teaserfigure} 
    \includegraphics[width=\textwidth]{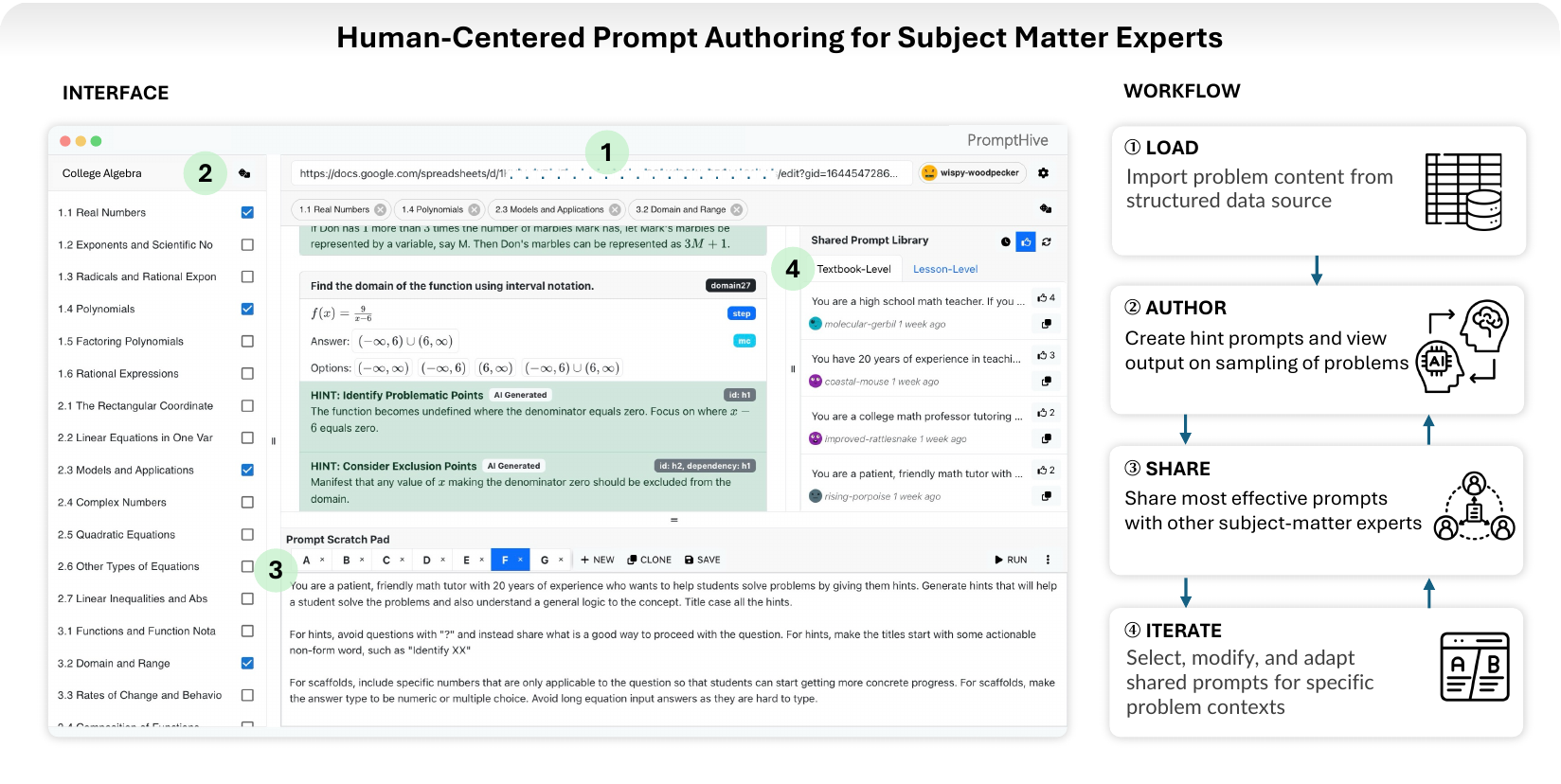} \caption{The \textbf{PromptHive} Interface \& Workflow: (1) \textbf{Load:} Import textbook lessons and problems by pasting a link to a content pool spreadsheet. (2) \textbf{Author:} Create hint prompts and view output on a variety of lessons and problems using dice-icon sampling buttons. (3) \textbf{Share:} Save effective prompt formulations to a shared library for other subject matter experts to clone, evaluate, and modify. (4) \textbf{Iterate:} Refine effective prompts in the scratchpad by editing and  comparing output using buttons labeled A, B, C, etc.} \Description{A screenshot of the PromptHive Interface, showing four workflow components: load, sample, evaluate, and share.} 
    \label{fig:teaser}
\end{teaserfigure}

\received{12 September 2024}

\maketitle

\section{Introduction}
As frontier Large Language Models (LLMs) push the boundaries of what computers can help humans create, the question of how to design authoring interfaces that effectively connect domain experts with the prompt engineering process becomes increasingly salient. With the right design \cite{tohidiCHI06}, such interfaces could enable experts to steer the output of LLMs toward content that better aligns with the nuances and needs of their domains, and transform the role of the subject matter expert from a \textit{producer} to a \textit{curator}—a competent and critical judge who \textit{instructs} the AI agent on what is needed, \textit{evaluates} the output, and \textit{iterates} on the instructions until the results are satisfactory. Instead of replacing human experts, these interfaces could help bridge human intelligence with machine intelligence to dramatically reduce the time and effort required to create content that adheres to expert tastes and standards.

To realize the \textit{producer-to-curator} shift and integrate domain expertise more closely into prompt engineering, we need authoring interfaces that: (i) deeply embed LLMs within existing expert workflows, augmenting content creation with carefully scaffolded interface support for prompt engineering; (ii) encourage experimentation on many prompt variations to systematically test the impact of changes in instructional wording on model output; (iii) offer mechanisms for curating prompt formulations that work well at various levels of abstraction; (iv) integrate generation into the publishing workflow. However, designing authoring interfaces that support experts across all four fronts is difficult as LLMs pose unique usability challenges tied to high metacognitive demands during prompt construction \cite{tankelevitch2024}, and users can struggle to get the models to integrate well with their existing workflow as even small perturbations such as adding a space at the end of a prompt can cause the LLM to change its output \cite{salinas2024butterfly}. For domain experts who aren't AI specialists, recent literature on prompt engineering has also highlighted how designing effective prompts can be surprisingly difficult for non-AI experts \cite{zamfirescu2023, brade2023promptify}.

In this work, we explore ways to enable domain expert-driven prompt engineering and answer empirical questions on how they permute and evolve prompts through the design of \textbf{PromptHive} (Figure \ref{fig:teaser}), an open-source collaborative prompt authoring interface that we apply to the content authoring workflow of an open-source adaptive tutoring system, OATutor \cite{pardos2023oatutor}, to support domain experts with writing prompts for generating hints to homework problems. PromptHive encourages rapid experimentation and systematic testing of alternative prompt variations via a set of randomization features for sampling problems from a spreadsheet, as well as buttons that pair prompt variations in the scratchpad with corresponding model output for easier comparison. This approach has been shown to be effective in a recent Human-AI authoring interface in a different context \cite{reza2024abscribe}. To support prompt curation, PromptHive features a shared library where users can save, clone, and upvote useful prompt formulations at two different levels of abstraction--the textbook-level and lesson-level. PromptHive also features a back-end logging engine that collects rich user interaction data on how prompts evolve as users create, modify and share them. 

To validate our design, we conducted two studies: (i) a three-stage user study with ten subject matter experts who had prior experience in manually authoring hints using the expert workflow augmented by PromptHive; (ii) a learning gain study with 358 learners, comparing the efficacy of hints generated using PromptHive with hints previously authored manually by subject matter experts. Our findings provide rich empirical data on how users collaboratively iterate on prompts to generate hints for an entire college-level algebra textbook, and show that users can successfully use PromptHive to create hints that are on par with those authored manually by subject matter experts who did not make use of generative AI, while cutting mental workload by \textit{more than half} and the time required from several months to just a few hours. The subject matter experts rated PromptHive as having excellent usability (SUS score = 89/100) and felt that it fostered trust in the performance of the LLM model (Figure \ref{fig:trust}). In this work, we contribute:
\begin{enumerate}
    \item The design and implementation of PromptHive, an open-source prompt-authoring interface that enables subject matter experts to collaboratively create prompts to generate hints to problems in an open-source adaptive tutoring system, dramatically reducing the time and effort needed for expert-driven hint creation.
    \item Results from two studies — a user study with ten mathematics subject matter experts and a learning gain study with 358 college students — demonstrating the efficacy of PromptHive and its advantages over manual content creation and curation.
    \item A tree-based backend logging engine that captures rich user interaction data on how subject matter experts permute and evolve prompts while using PromptHive.
\end{enumerate}


\section{Related Work}

We review the literature in HCI and Generative AI to explicate the need for integrating subject matter expertise into prompt engineering from both ethical and practical standpoints. We also highlight the challenges associated with designing human-centered interface support for prompt engineering and examine existing systems for collaborative prompt engineering to situate and distinguish PromptHive from other tools.

\subsection{Integrating Subject Matter Expertise into Prompt Engineering}

Placing subject matter experts in the driver's seat of prompt engineering is crucial as they possess the necessary judgement to evaluate the output of LLMs in their domain \cite{sambasivan2022deskilling, kumar2022data}. It is also the case that software engineering teams may not be representative of the users and their demographics \cite{adams2020diversity}, and therefore may be less able to produce prompts that address a variety of needs and perspectives. Domain expert-driven prompt engineering is also important because of the wide-ranging and lasting impact LLMs are having across many, many domains including art \cite{ko2023large, zhou2024generative}, medicine \cite{zhang2023generative, mesko2023imperative}, law \cite{harasta2024cannot, fei2023lawbench}, and education \cite{kumar2023impact, pardos2024chatgpt}. Unlocking the potential of pretrained generative models hinges on aligning them with human intentions \cite{wang2024one}. Researchers, organizations, policy makers, and society at large will need to grapple with the question of how to best involve human experts in AI-intensive workflows. Preliminary studies have shown the potential benefits of expert involvement in the application of LLMs. For example, Kumar et al. found that an instructor-tuned LLM significantly boosted student interactions with a chatbot compared to plain ChatGPT as a baseline \cite{kumar2023impact}, and Wang et al. extended the Mixture-of-Experts Paradigm to prompt optimization, demonstrating how breaking up a problem space into sub-regions controlled by specialized human experts can help with prompt optimization \cite{wang2024one}.

In this work, we contribute an open-source system that exemplifies how to effectively integrate subject matter expertise in the context of educational content creation and contribute to the ongoing discourse within the HCI and Human-Centered AI communities on retaining human control while increasing automation \cite{shneiderman2022}. In contrast to Sheridan and Verplank's characterization of automation and human control as a unidimensional spectrum \cite{sheridan1978human, shneiderman2020human}, we adopt Shneiderman's two-dimensional HCAI framework \cite{shneiderman2020human} and explore ways to increase automation \textit{without} encroaching upon subject matter experts' sense of control and trust over the content. 

\subsection{Usability Challenges of Designing Prompt Engineering Interfaces}

At the same time, as noted by Tankelevitch et al., the same unique properties that make LLMs powerful, such as their flexibility across multiple input/output spaces and generality across tasks, pose usability challenges for the design of human-centered Generative AI systems due to the high metacognitive demands of prompting \cite{tankelevitch2024}. These usability challenges mean that subject matter experts need interface support to carefully scaffold the task of prompt writing, so that they can focus on the core value they bring to the table, i.e., instructing the LLM on what is needed, evaluating the output, and iterating on the instructions until the output is satisfactory. Such scaffolding must take into account how unpredictable generative models can be and make experimentation and iteration a core part of the prompt design process. Salinas and Morstatter describe the butterfly effect of prompt alterations – even minor and seemingly arbitrary changes such as adding a space at the end of a prompt, ending with "Thank you," or promising the LLM a tip, can alter the LLM output \cite{salinas2024butterfly}. 

In PromptHive, we design and validate explicit interface support for iterating on multiple prompt variations in the form of prompt-output buttons that pair prompts with the generated output in a scratchpad interface to allow experts to qualitatively compare the impact of changes to model output. We leverage the latest advancements in LLMs to deal with model unpredictability, such as the ability to reliably adhere to specific output schemas in GPT4o, self-consistency prompting \cite{wang2022self}, and multimodal capabilities that take into account images in problem content, to effectively integrate PromptHive within a real-world expert workflow for generating hints in a production tutoring system \cite{pardos2023learning}.




\subsection{Designing Effective Content Authoring Tools for Intelligent Tutoring Systems}

Intelligent Tutoring Systems (ITS) have historically required highly time consuming authoring processes. Early iterations of authoring tools required 200-300 hours of use to produce a single hour of instructional content and required experience with programmatic interfaces \cite{aleven2006cognitive}. Development of the Cognitive Tutor Authoring Tools (CTAT) quickly reduced this time to 50-100 hours to produce one hour of instructional content \cite{weitekamp2020interaction}. CTAT provided content authors with a GUI which allowed for content authoring with minimal knowledge of programming and coding skills. The GUI allowed for content creation to take place with the respective interface, which greatly supported independent content creation processes \cite{aleven2006cognitive}.

Following the example of CTAT, many future ITS and ITS-like systems supported their content creation ecosystems with sets of builder tools that were focused on providing content authors an easy-to-use interface, usually in the form of a GUI. One example of such tools is that of the ASSISTments builder. While not fully an ITS, ASSISTments is an adaptive tutor that places the instructor in an important role within its system \cite{heffernan2014assistments}. Supporting such a design philosophy, the ASSISTments Builder was created to simplify the content authoring process for teachers and instructors \cite{razzaq2009assistment}. To accomplish this, the builder provided easy integration of problem-help features such as hints and scaffolds, while also allowing skill mapping of knowledge components. Furthermore, it managed to match the lower end of CTAT’s 50 hour estimate of development for one hour’s worth of instructional content despite using a GUI instead of a programmatic interface \cite{razzaq2009assistment}. The ASSISTments Builder also supported problem variabilization directly through its GUI, allowing for easier variation of problem content \cite{razzaq2009assistment}. 

While GUIs have provided many benefits for content authoring, there are still many challenges that accompany them. With the large number of tutoring systems available, it can be difficult for teachers to have to learn a new content authoring environment every time they wish to utilize a new tutor \cite{weitekamp2020interaction}. Furthermore, creating problems with more complicated structures may not just be difficult, but in some GUIs may not even be possible \cite{razzaq2009assistment}. This has resulted in difficulties for new systems to balance between efficient and easy content creation while also supporting more complex content creation. When compared to the GUI of the ASSISTments builder, the programmatic spreadsheet-based authoring interface used in OATutor  did not result in any significant differences regarding time taken for content authoring, but showcased higher accuracy when curating content, albeit with a considerably lower usability score \cite{sheel2024comparing}.

These challenges underscore the need for the next generation of LLM-infused content authoring systems to integrate seamlessly with \textit{existing} expert workflows, ensuring that AI adapts to the needs of users, rather than the reverse. In this work, we demonstrate how systems like PromptHive can successfully augment complex workflows involving multiple experts with minimal disruption.

\subsection{Generative AI in Tutoring Systems}

Recent years have seen widespread implementation of generative AI tools into digital technologies. LLM improvements have had a significant influence on tutoring systems and educational contexts. Evaluations of ChatGPT’s decimal skills indicated that it can respond accurately to conceptual questions but struggles more with respect to number lines and decimal point problems \cite{nguyen2023evaluating}. Furthermore, when examining student answers, ChatGPT accurately assessed the correctness of seventy-five percent of them, while generating feedback that was similar to that of instructors. ChatGPT's ability to produce informative worked solutions proved effective for learning in Algebra and Statistics subjects \cite{pardos2024chatgpt}. In relation to higher education, across 53 studies, with 114 question sets totaling to over 49,000 multiple choice questions (MCQs), ChatGPT 4.0 answered 75.5\% of all MCQs on said problem-sets, receiving a passing score on the majority of them \cite{newton2023chatgpt}. It has also been shown to effectively evaluate the quality of learnersourced MCQ distractors \cite{moore2023assessing}.

LLMs have seen usage with respect to tutor question quality evaluation. Generative Students is a prompt architecture that utilizes LLMs to simulate student profiles for the purpose of simulating believable MCQ answers \cite{lu2024generative}. Generative Students demonstrated a high correlation between how real students responded to the sample question and how the simulated students did. Furthermore, the system demonstrated that an instructor could improve their question quality using these simulated students. A similar approach was also used to learn question difficulty parameters using LLM-Respondents, thereby allowing for appropriate quality tutoring questions to be selected \cite{liu2024leveraging}. Beyond simulated students, LLM tutoring use also has taken the form of chatbots. An ASSISTments integrated chatbot utilizing GPT 4.0, while not providing statistically significant learning gains, increased the positivity of students' attitudes, even though they displayed a lower confidence of solving a similar problem after the chatbot help intervention \cite{cheng2024facilitating}.

LLM-generated questions themselves have also been evaluated. When compared against questions from a published Creative Commons textbook, there were no statistically significant differences between the difficulty of algebra textbook questions and similar ones generated by ChatGPT \cite{bhandari2024evaluating}. These results indicate that ChatGPT is capable of producing algebra problems of similar quality to those from textbooks.





\section{Designing PromptHive}
In this section, we describe the expert workflow that PromptHive is designed to support, the core design requirements, the development process, and key interface elements.

\subsection{Understanding the Expert Workflow}
\begin{figure*}
  \centering
  \includegraphics[width=\linewidth]{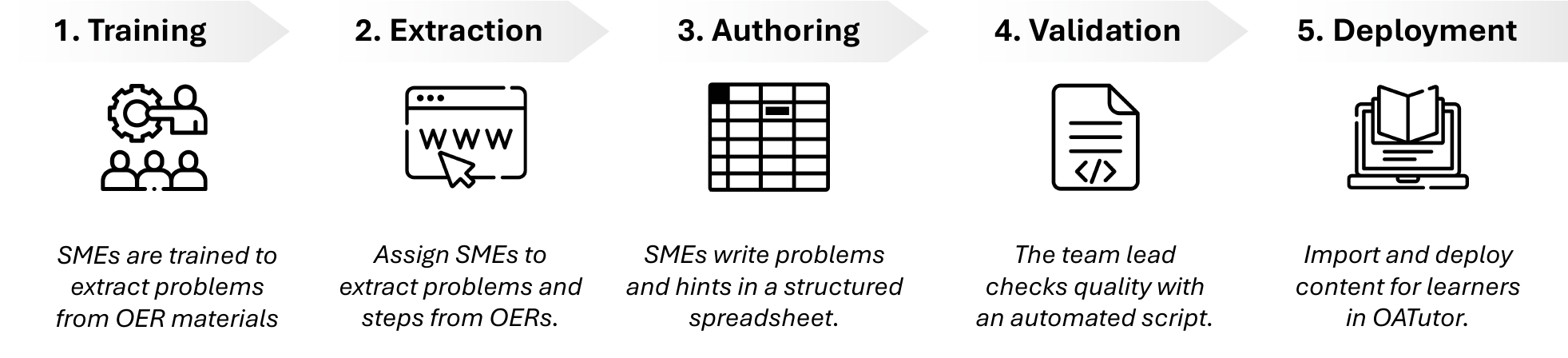}
  \caption{The subject matter expert workflow for manually authoring content in OATutor.}
  \label{fig:workflow}
\end{figure*}

To understand how generative AI could support educational content creation, we explore the details of the existing expert workflow within OATutor. Open Adaptive Tutor (OATutor) was selected for its Creative Commons content library, rapid experimentation capabilities, structured json-based content format, and documentation of the expert-authoring of its content \cite{anastasopoulos2023exploring}. Tutoring systems based on ITS principles have proven effective for learning \cite{pane2014effectiveness}, as has OATutor in the subjects of Algebra and Statistics \cite{pardos2024chatgpt}. The lead author conducted a series of in-depth conversations with the head of OATutor's content team, who oversees the training of subject matter experts (SMEs) in content authoring. As outlined in Figure \ref{fig:workflow}, the expert workflow that has previously been employed in OATutor followed a five-step process:

\begin{enumerate}
    \item \textbf{Training:} SMEs with prior tutoring experience in different subject areas, including mathematics, physics, or chemistry were trained by the content team lead on how to extract problems from open educational resources (OERs) such as free textbooks from the OpenStax project \cite{stafford2018openstax}. An introductory Canvas course was provided to the SMEs to prepare them accordingly.
    \item \textbf{Extraction:} Then, SMEs were assigned specific chapters or topics from the OER materials and tasked with curating problems, breaking them into smaller, manageable steps.
    \item \textbf{Authoring:} The problems were entered into a structured spreadsheet. Each row represented a problem and its associated steps, and SMEs authored two types of hints for each step: plain hints without answers and scaffolded hints that provided step-by-step solutions. The spreadsheet required adherence to strict formatting rules, including custom math formula formats, and this entire authoring process was done manually. 
    \item \textbf{Validation:} Senior content team members conducted quality checks using an automated validation script to ensure adherence to the correct structure and format in addition to manual checks for correctness and aspects such as spelling and grammar.
    \item \textbf{Deployment:} Once validated, the content was imported into OATutor, making it available for students. Once on the system, feedback could be received regarding the problems, which could lead to further iterations and fixes.
\end{enumerate}

\subsection{Eliciting Design Requirements}
To effectively integrate generative AI into this expert workflow, we surveyed frameworks for Human-AI collaboration and prompt-authoring interfaces for LLMs to derive an overarching design requirement (R0) and five supporting design requirements (R1-5) for PromptHive:

\begin{itemize}
    \item \textbf{Control (R0):} Ensure SMEs retain control while leveraging automation. Drawing on the two-dimensional Human-Centered AI (HCAI) framework by Shneiderman \cite{shneiderman2020human}, we aimed to automate the content generation process as much as possible without diminishing SMEs' sense of control. To achieve this, we provided interface tools that allowed SMEs to easily guide content generation and oversee quality, while automation reduced the effort required to produce educational materials.
    \item \textbf{Integration (R1):}  Seamlessly integrate AI support within the existing expert workflow. The system needs to read and write content in the same format used by SMEs and preserve human oversight over the AI-generated content. Integration with current processes is crucial to ensure smooth adoption without disrupting established workflows.

    \item \textbf{Simplicity (R2):} Given the usability challenges associated with human-centered generative AI, particularly the high metacognitive demands of prompt engineering \cite{tankelevitch2024}, a key requirement is to minimize cognitive load by scaffolding the prompt-writing process. The system should enable SMEs to focus on content quality — such as instructional clarity and hint structure — rather than dealing with technical complexities like API management or formatting rules.

    \item \textbf{Trust (R3):} Building trust in AI is a central challenge identified in the explainable AI (XAI) literature \cite{arrieta2020explainable}. Therefore, a core requirement is to foster trust in the system by providing SMEs with tools to evaluate and validate AI-generated content thoroughly, ensuring alignment with their educational goals and quality standards.

    \item \textbf{Iteration (R4):} Given the inherent unpredictability of LLM outputs, experimentation has been recognized as a critical requirement for effective prompt-authoring \cite{reza2024abscribe, kim2023cells}. PromptHive must support rapid iteration on multiple prompt variations, enabling SMEs to compare outputs and refine prompts efficiently to achieve optimal results.

    \item \textbf{Collaboration (R5):} Finally, given that the existing expert workflow involved multiple content team members, in alignment with the social paradigm of prompt design \cite{wang2024wordflow}, we seek to support collaborative prompt engineering to write and refine effective formulations together.
\end{itemize}

\subsection{Brainstorming \& Design Review Sessions}

To transform these design requirements into concrete interface support for subject matter experts in PromptHive, the lead author held weekly brainstorming and design review sessions with the content team lead over a span of six months and engaged in rapid prototyping, transitioning from low-fidelity sketches to the fully-functional open-source system that this paper contributes. The design process involved three key stages: (i) Paper-Prototyping: Low-fidelity sketches explored ways for SMEs to integrate human-authored content with generative AI and rapidly iterate on hints; (ii) Cognitive Walkthroughs: Informal sessions were held to refine the paper prototypes and ensure the designs were intuitive; (iii) Web-Based Prototyping: High-fidelity prototypes were built based on the refined designs. There was significant overlap and back-and-forth iteration throughout these stages, with feedback from walkthroughs informing subsequent changes.

\subsection{Developing the PromptHive Interface}
\label{sec:prompthive_workflow}

\begin{figure*}
  \centering
  \includegraphics[width=\linewidth]{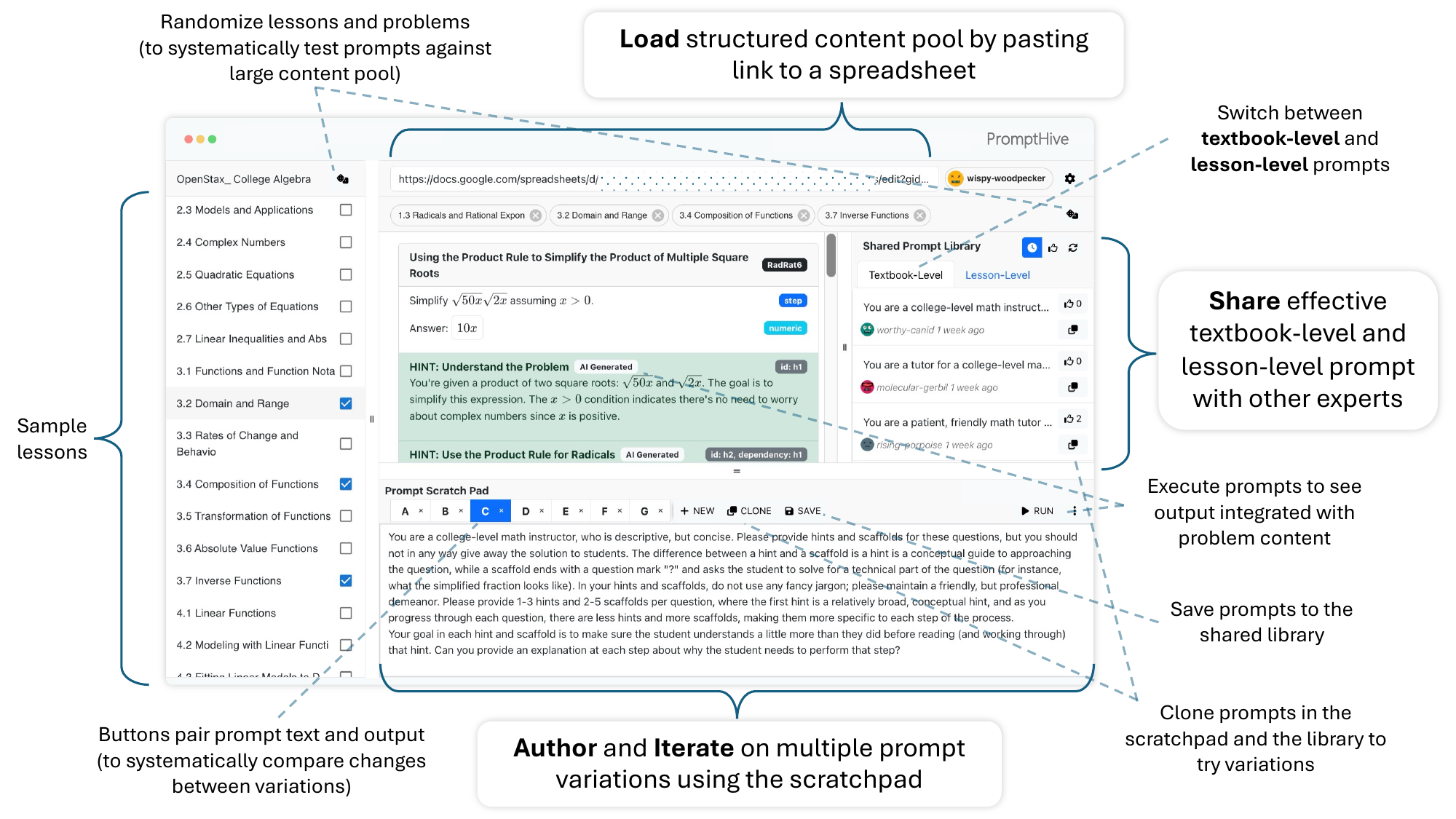}
  \caption{How the interface elements in PromptHive map to the 4-stage, 2-level workflow.}
  \label{fig:interface}
\end{figure*}

This iterative process led to the development of an interface that supports a 4-stage prompt-authoring workflow at two levels of abstraction, as described in Figure \ref{fig:interface}:

\begin{enumerate}
    \item \textbf{Load:} To fulfill R1 and enable seamless integration between human-authored and AI-generated content, we implemented a loading mechanism in PromptHive that allows SMEs to import content from structured data sources, such as an OATutor spreadsheet, by pasting a link.
    \item \textbf{Author:} Users can author hint prompts and view its output on a sampling of problems. To address R2 and R3, and ensure SMEs can thoroughly test prompts against a variety of problems and lessons even when the content pool is large, we introduced randomization buttons. These allow users to systematically sample problems for more comprehensive testing.
    \item \textbf{Share:} To fulfill R5 and facilitate collaborative iteration among teams of SMEs, we developed the Shared Prompt Library. This feature allows SMEs to curate effective prompt formulations, enabling others to clone, test, and upvote prompts that perform well during evaluation.
    \item \textbf{Iterate:} To fulfill R4 and support rapid iteration on prompt variations, the "Prompt Scratch Pad" interface enables SMEs to edit and experiment with multiple prompts in parallel. Users can clone prompts to test variations, compare generated outputs side by side, and quickly evaluate how changes impact the content by clicking on buttons that pair each prompt with its corresponding output.
\end{enumerate}
These processes are more cyclical than linear, meaning that sampling, evaluation, and sharing can occur at any stage of the prompt crafting process, allowing for frequent back-and-forth iterations between content team members. Prompt authoring happens at two levels of abstraction:
\begin{enumerate}
    \item \textbf{Textbook Level:} Users create prompts that generate hints applicable across an entire textbook, ensuring broad coverage.
    \item \textbf{Lesson Level:} Prompts are then refined to address specific lessons, ensuring greater specificity and alignment with particular learning objectives.
\end{enumerate}

In the next two sections, we describe methodological details and results of two studies that we conducted to refine and validate our design: (i) an in-depth, three part study with ten subject-matter experts to refine and validate the interface design and write prompts for generating hints for an entire college-level algebra textbook; (ii) a learning gain study with 225 learners to assess the learning impact of hints generated using PromptHive compared to human-authored hints created previously by SMEs during previous years without AI help. Both study protocols were approved by the UC Berkeley Committee for the Protection of Human Subjects under IRB Protocol 2024-05-17449.

\section{Study 1: User Evaluation with Subject Matter Experts}

To evaluate how subject matter experts use PromptHive to collaboratively author prompts, we designed a three-part user study involving pre-interviews, collaborative prompt-writing sessions, and post-interviews with experts in math.

\subsection{Research Questions}

We sought to answer the following research questions through this study:
\begin{itemize}
    \item RQ1: What do SMEs perceive as the most \textbf{time-consuming} tasks in manual content authoring, and where can AI offer the greatest assistance through PromptHive?
    
    \item RQ2: How do subject-matter experts with prior experience in manual content authoring perceive the \textbf{trustworthiness} and \textbf{usability} of PromptHive?

    \item RQ3: Do SMEs feel they can retain \textbf{control} over the hint generation process when using PromptHive?

     \item RQ4: How does the subjective \textbf{cognitive workload} of SMEs using PromptHive compare to the manual content generation workflow?
    
    \item RQ5: How do prompts \textbf{evolve} as subject-matter experts refine them through individual iterations and collaboration with other experts?
\end{itemize}

\subsection{Participants}
We recruited 10 subject-matter experts (6 women, 4 men) in mathematics tutoring, aged 18-24, who were screened for having prior experience in manually authoring content for OATutor. These participants were either 3rd- or 4th-year undergraduate students or recent graduates, with backgrounds in math-intensive fields such as Data Science, Computer Science, Applied Mathematics, Economics, and Electrical Engineering. All participants except P9 reported having experience tutoring math and other subjects beyond their content authoring work for OATutor. However, P9 had extensive experience authoring content for OATutor, contributing around 500 problems to the platform. Each subject-matter expert was compensated with \$150 USD for participating in the study. Table \ref{table:participants} summarizes their backgrounds and expertise. Participants had some prior exposure to LLMs, primarily through ChatGPT, but were not advanced users with knowledge of APIs or other technical aspects of prompt engineering.

\begin{table}[h!]
\centering
\small 
\renewcommand{\arraystretch}{1.2} 
\begin{tabular}{l p{3.5cm} p{3.5cm} p{2cm} p{3.5cm}} 
\toprule
\textbf{ID} & \multicolumn{2}{c}{\textbf{Educational \& Teaching  Background}} & \multicolumn{2}{c}{\textbf{OATutor Authoring Experience}} \\ 
\cmidrule(lr){2-3} \cmidrule(lr){4-5}
            & \textbf{College Major} & \textbf{Teaching Experience} & \textbf{\# of Problems} & \textbf{Problem Subject Areas} \\ \midrule
P1          & Applied Mathematics & TA in Data Science                                          & 100                         & Calculus                        \\ 
P2          & Legal Studies, Economics              & University tutor for Calculus                               & 150                         & Physics, Calculus, Algebra                          \\ 
P3          & Computer Science                      & TA for Introductory CS courses               & 75                          & Chemistry, Algebra,  Physics \\ 
P4          & Computer Science                      & High school math tutor, debate coach (middle school - college) & 100                         & Calculus, Statistics                                                \\ 
P5          & Mathematics  & 7th-Grade Math Teacher, also tutored AP Statistics and College-level Algebra      & 150                   & Algebra\\ 
P6          & Computer Science, Data Science        & Tutor for statistics, math, and programming                 & 200                         & Statistics, Algebra                                 \\ 
P7          & Data Science              & Tutor for high-school Math                               & 200                         & Algebra                          \\ 
P8          & Computer Science, Cognitive Science   & Instructor and TA for Computer Security, Intro to CS tutor  & 325                         & Statistics, Algebra \\ 
P9          & Data Science, Cognitive Science       & No tutoring beyond OATutor                          & 500                         & Physics, Calculus \\ 
P10         & Electrical Engineering, Computer Science & Tutor for College-level CS                            & 120                         & Mathematics and Physics           \\ 
\bottomrule
\end{tabular}
\caption{Background Summary of Subject Matter Experts.}
\label{table:participants}
\end{table}

\subsection{Procedure \& Tasks}
In addition to completing a short demographics survey, participants took part in three study sessions, all done only via Zoom video-conferencing:

\begin{itemize}
    \item \textbf{30-minute Pre-Interviews:} These sessions were completed individually and began with participants sharing their prior experience with manually authoring content for OATutor. They were then asked to reflect on this experience and complete the NASA-TLX instrument for measuring perceived workload. Following this, participants were given access to an early version of the PromptHive interface and asked to complete a series of steps following the 4-stage workflow outlined in Section \ref{sec:prompthive_workflow}. During this think-aloud session, the researcher noted any usability issues that emerged, which were then addressed before the collaborative prompt-writing sessions described next. 
    \item \textbf{1.5-hour Collaborative Prompt Writing:} These were conducted in groups, with participants choosing between two available time slots. This two-session approach allowed us to simulate both synchronous and asynchronous collaborative scenarios, as prompts written during the first session were made asynchronously available to participants in the second session. The sessions began with participants receiving an overview of how the PromptHive system works, along with a document outlining the necessary instructions and basic prompt-writing strategies adapted from OpenAI's guide to prompt engineering \cite{promptguide}. We excluded technical details and focused on two core strategies: testing changes systematically and writing clear instructions. Participants first worked individually to create their best textbook-level prompts. Next, they were randomly and evenly assigned contiguous groups of lessons (most participants received six lessons and one received five, dividing 59 lessons across 10 participants) to test and refine both their own and others' textbook-level prompts. They could upvote effective prompts and adapt them at the lesson level if the textbook-level versions did not work well as-is. By the end of the session, each participant contributed a lesson-level prompt to the shared library, collectively covering the entire college-level algebra textbook.
    \item \textbf{30-minute Post-Interviews:} After participants completed the collaborative prompt writing sessions, they scheduled individual post-interviews to share their experience with using PromptHive. They reflected on their experience for generating hints using PromptHive and completed the NASA-TLX \cite{hart2006nasa} instrument for the automated workflow. They also completed the System-Usability Scale (SUS) \cite{lewis2018system} and the XAI Trust scale adapted from Hoffman et al. \cite{hoffman2023measures} questionnaire to share how usable the system felt and whether they trusted using it for the purpose of generating educational content. See Appendix \ref{appendix:trust} for the wording of the 8 items in the trust scale. Finally, they participated in a brief semi-structured user interview to unpack their experience with PromptHive.
\end{itemize}



\subsection{Materials}
We used publicly available Creative Commons Algebra content from Open Adaptive Tutor (OATutor) \cite{oatutorgithub}, an open-source adaptive tutoring system based on ITS-principles \cite{pardos2023oatutor}. This content pool contains materials from the \textit{OpenStax College Algebra 2e textbook} \cite{openstaxalgebra}; we contacted the OATutor team to share a copy of the spreadsheet they used during manual authoring. This spreadsheet was ideal because it contained pre-existing human-authored hints and scaffolds that we could compare with those generated using PromptHive, as we do in the second study which focuses on learning gain. We chose mathematics because domain expertise could be particularly valuable for this subject area as LLMs are known to struggle with mathematical reasoning \cite{singh2024exposing}, making subject matter expert involvement valuable. It is also an area where responsible applications of generative AI could have immense positive impact on students \cite{bin2022artificial}, and recent literature suggests strategies for reducing hallucinations, such as self-consistency \cite{wang2022self, pardos2024chatgpt}, an approach that we adopt later for the second study. 

\subsection{Analysis}
Our data consisted of interview transcripts and NASA-TLX ratings from the pre- and post-interviews, SUS and XAI Trust ratings from the post-interviews, the textbook-level and lesson-level problems that participants wrote during the collaborative sessions, and the detailed prompt iteration JSON logs captured by the logging engine. We analyzed the qualitative interview data using refexive thematic analysis \cite{braun2019reflecting} through an inductive-deductive lens, using the rich theory on Human-AI collaboration and recent literature on prompt-engineering interface design as a pre-existing code that guided our interpretations. For the NASA-TLX scores and SUS ratings, we used the standard procedures for calculating scores. 

\section{Study 1 Results}

\begin{figure}[h!]  
  \centering
  \includegraphics[width=0.8\textwidth]{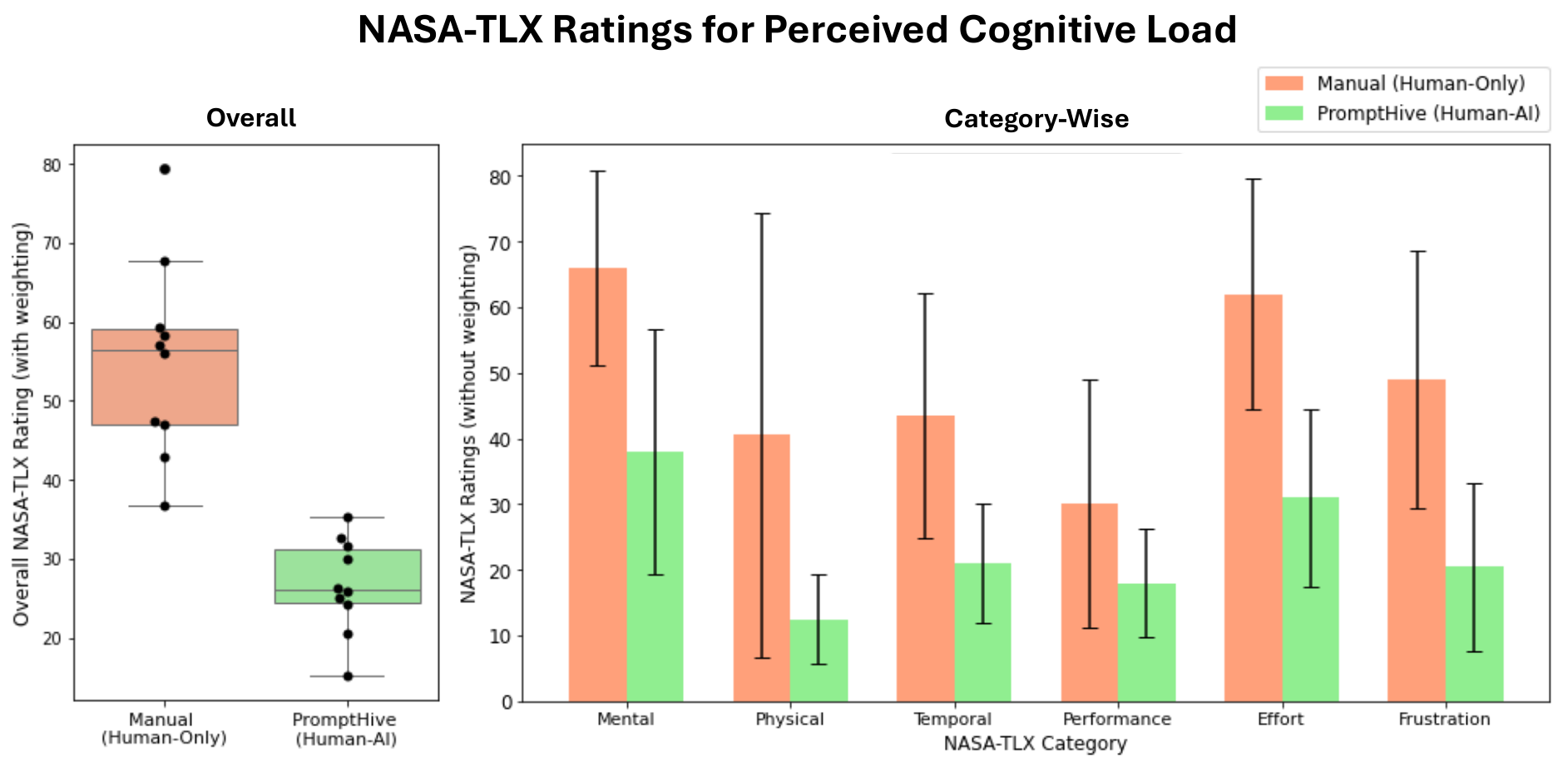}  
  \caption{NASA-TLX ratings for the content authoring workflow in PromptHive versus manual.}
  \label{fig:nasatlx}
\end{figure}

\begin{figure*}
  \centering
  \includegraphics[width=0.8\linewidth]{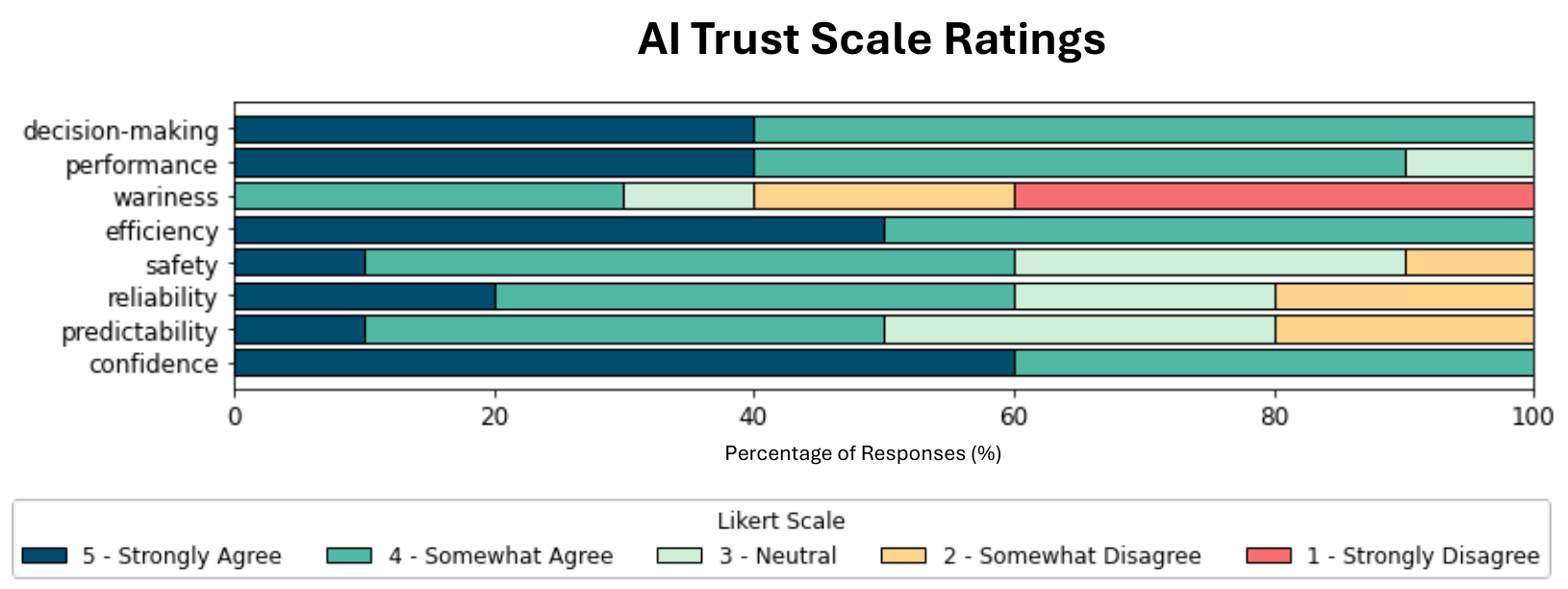}
  \caption{Participant rating distribution for the Explainable AI (XAI) Trust scale adapted from Hoffman et al. \cite{hoffman2023measures}} Note that for wariness, higher disagreement is desirable. For other items, higher agreement is desirable. 
  \label{fig:trust}
\end{figure*}

The overall response to PromptHive was largely positive, with participants rating the system as highly usable, achieving an SUS score of 89/100. There was a significant reduction in NASA-TLX ratings for perceived cognitive workload, dropping from 55.17 to 26.73 out of 100 compared to the manual hint authoring workflow (see Figure \ref{fig:nasatlx}). Regarding trust in the AI system, most participants strongly or somewhat agreed that the system felt trustworthy, without causing any wariness, based on the validated measures from the XAI trust scale. See Figure \ref{fig:trust} for the distribution of responses and Appendix \ref{appendix:trust} for the wording of the scale items adapted from \cite{hoffman2023measures}.

\subsection{Unpacking the AI-assisted Hint Authoring Experience in PromptHive}

\begin{quote}
   \textit{ ``Honestly, I’m just amazed by how [PromptHive] works. I’ve spent the last six months making hints, so seeing this in action and knowing that this is what other people will use is pretty incredible. I’m excited to see where it goes.''} -- P10
\end{quote}

During the pre- and post-interviews, participants shared their experience with the PromptHive system and commented on how it compares with the manual workflow and their prior experiences with AI. We grouped the findings from their experiences into six themes (F1-6) described below. 

\textbf{F1: Hint Authoring is Most Time-Consuming and Can Benefit from AI Assistance}

To answer RQ1, we asked participants during the pre-interviews which parts of the manual hint authoring workflow they found most time-consuming. We posed this question \textit{before} introducing participants to the PromptHive system to avoid biasing their responses. Multiple participants (P1, P3, P5, P7, P9) pointed to authoring hints (especially those with scaffolded answers), as the most time consuming, in alignment with our design decision to target that aspect of the content authoring process using PromptHive. For example, P5 shared that, ``the most time-consuming part is how to scaffold each step to make it clear and short without including too many words,'' while P3 mentioned, ``each problem has only one statement, but for the hints, there could be 10 or more.'' 

Regarding why scaffolded hint generation was time-consuming, P9 discussed the decision-making process involved in determining the appropriate level of detail for hints, such as whether to explain high-level concepts or break down smaller calculations. They questioned, ``If there’s a problem that uses the mass-times-density equation, do I need to just say, ‘Use a density equation and do the calculation with these inputs,' or go further down to something like, `What’s three times five?''' P10 shared additional considerations when authoring hints, noting that it takes time to ``...chain the hints and ensure they flow together.''

In contrast, post-interviews revealed a clear consensus among participants that PromptHive significantly reduced the time spent creating hints compared to the manual workflow. Participants noted that they could generate multiple hints and scaffolds in the time it previously took to create just one manually. This efficiency was particularly important when handling large volumes of problems. For instance, P8 remarked, ``I was able to create a prompt to generate hints and scaffolds for 15 to 20 questions, whereas before, it would take me at least five, sometimes ten minutes to do just one.'' P10 added that PromptHive felt “super fast, super intuitive,” and noted, ``Anyone who’s done it manually and then tried the AI system would probably agree that it’s much more fun and intuitive... much more enjoyable than just typing it all out.''

\textbf{F2: PromptHive Integrates AI into Expert Workflow More Closely Compared to ChatGPT}

Comparing PromptHive to using ChatGPT, P1 mentioned ``I also tried, like, ChatGPT, but it always outputs the wrong answer. At least when I was working with this system, the answers were all reliable. I’m curious – how could you generate the result with such a high percentage of correct answers compared to GPT?''. Commenting on the tighter integration with the existing expert workflow, P4 noted that PromptHive is ``a lot faster because you don't have to type the question out into ChatGPT, and then copy, paste everything onto the spreadsheet. And the formatting for the spreadsheet was also like, difficult to do, because in it you had to write six or seven rows for each question. So I think the system is a lot easier to use. I would recommend over manually putting things into ChatGPT.''

\textbf{F3: Participants felt that the AI-assisted workflow in PromptHive led to more consistent hints}

Several participants (P1, P7, P8 and P9) felt that the AI hints were more consistent than the human-only hints because of the variability in style or structure of the hints across different subject matter experts. Sharing reasons for the variability they observed, P1 emphasized that different members had different teaching styles. P7 felt the AI workflow was more uniform and less prone to error compared to the manual workflow. Pointing to a different reason for the observed inconsistency, P4 mentioned that the quality of hints would vary a lot, and ``sometimes you could see the effort decreasing from like the first question to like. So, yeah, I think it would vary, yeah'', as people got tired.

\textbf{F4: Participants felt that they mostly retained control over Hint Generation}

Many participants (P2, 6, 8, 9, 10) reported that they strongly felt they could steer the output of the LLM and apply their prior subject matter expertise in authoring prompts and evaluating the model output. For example, P6 was able to make hints simpler or complex by altering the student age: ``I tried asking the AI to explain to a 7-year-old, and the output was simple. When I changed it to a 10-year-old, it became more complex.'' P6 removed redundancy by instructing the LLM to not repeat hints and limit the number of scaffolded hints, like P9. P8 mentioned that they could eventually get the AI to follow instructions after iterating on the prompt a few times: ``even if there are times where they are acting a little bit, you know, a little bit weird, like with capitalization issues or the tone is not right, I can just instantly update it. And then, usually, after two or three tries, I can get like, a version that I think would be even better than something that I would write myself.''

Other participants felt less strongly about their ability to control the AI output, e.g., P4 mentioned that while the AI model felt ``steerable to an extent'', it could sometimes be ``stubborn'' and difficult to steer for general textbook level prompts but that it felt easier to steer prompts at the lesson level. P5 felt that sometimes they struggled to get the AI model to adhere to multiple opposing instructions, especially when the prompts were complex, e.g., in trying to simplify the hints for a younger audience, the model would inadvertently give responses that felt too long. 

\textbf{F5: Subject Matter Experts did not feel replaced by the AI workflow}

Reflecting on the ongoing discussions surrounding how generative AI will affect education, P8 said ``I think when I hear AI and LLMs and education right now, I think of a talk that I just attended that was just very extremely pro-LLM. And I think we do need to be a little careful about how we inject LLMs into education. But I think having this idea where the LLM isn't creating the questions, the questions are already there. That is good to me. But then also, beyond that, like your job is to generate prompts...at the end of the day, having a prompt engineer who still goes through the process of needing to understand how to guide a student, I think that that still can be very beneficial, versus just like a solely AI based thing.'' When asked if they felt replaced by the AI model, they said ''I don't think the AI system can replace me. I think my role in terms of telling the AI what to do, that involvement is really important.'' P6 felt that more complex problems could benefit from increased oversight of subject matter experts.  

\textbf{F6: Collaborative authoring can help Subject Matter Experts be more creative with prompts}

\begin{figure*}
  \centering
  \includegraphics[width=\linewidth]{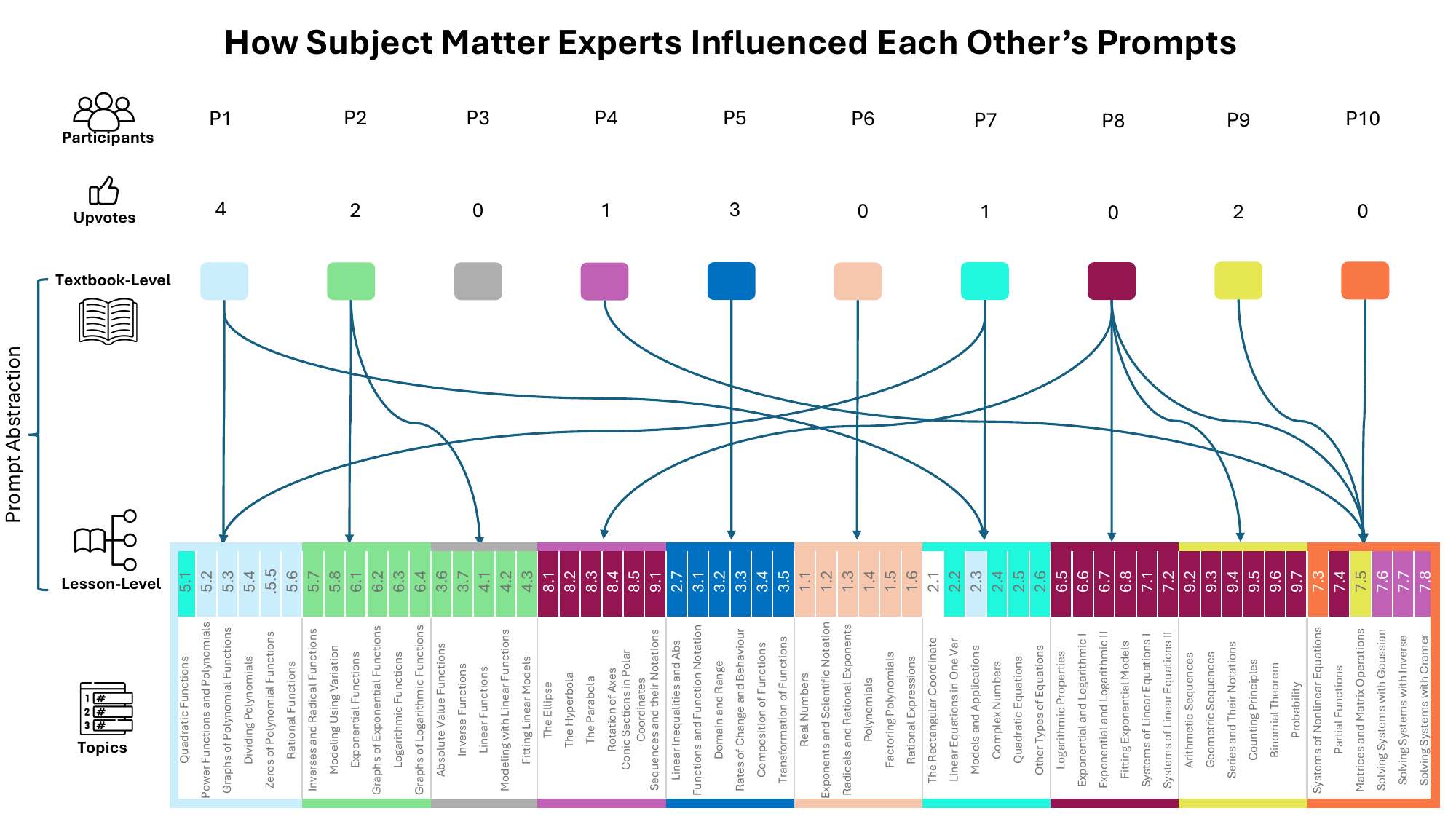}
  \caption{Participants' influence on each other's prompts. Outer border colors for lessons correspond to lessons assigned to participants, with the same color representing the same participant. The fill color of lesson numbers and the arrows indicate the textbook-level source for the lesson prompts.}
  \label{fig:collaborate}
\end{figure*}

The collaborative aspect of the AI assisted workflow in PromptHive was well-received by many participants, who felt that sharing ideas led to a richer variety of more creative prompts. P10 said ``seeing other people's prompts gave me ideas that I didn't even think to ask for'' and P8 felt ``the collaborative aspect was wonderful. Seeing how others structured their prompts helped me brainstorm and improve my own.'' Comments from P5 indicated that exposure to prompts from others helped them think beyond their own experience as a middle school teacher -- ``I remember seeing prompts asked for hints to have a positive tone, stuff like that. I think it inspired me to think about that because currently I'm a middle school teacher, so I only think about making things short so people will look at it. When I saw another prompt which said you are a tutor in college, you are a professor, that made me think about the main population of students this model is targeting.'' Figure \ref{fig:collaborate} provides details on how participants influenced each others' lesson-level prompts.

\subsection{Exploring How Subject Matter Experts Iterate on Prompts}

\begin{figure}[!tbp]
  \centering
  \begin{minipage}[b]{0.49\textwidth}
    \includegraphics[width=\textwidth]{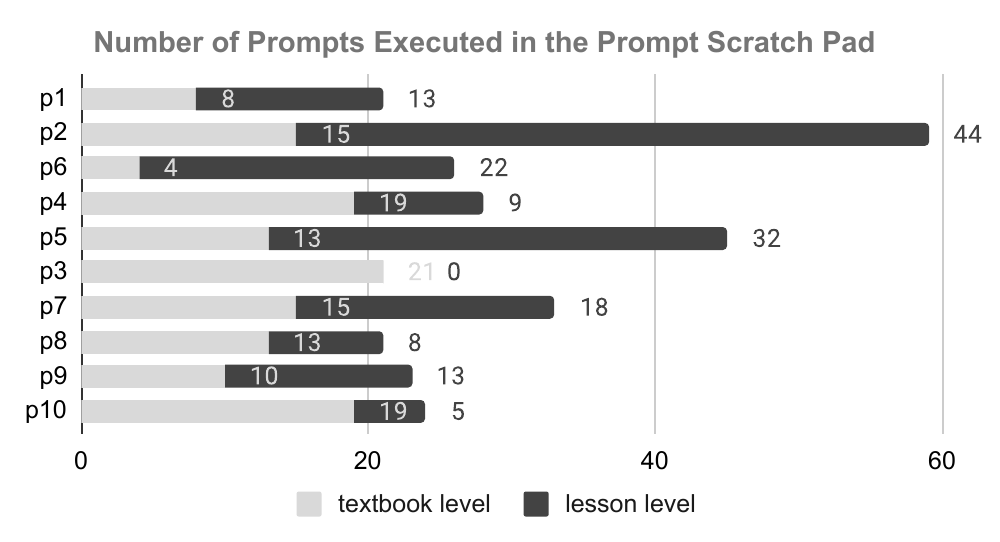}
    \caption{The number of times participants \textbf{executed} prompts.}
    \label{fig:executions} 
  \end{minipage}
  \hfill
  \begin{minipage}[b]{0.49\textwidth}
    \includegraphics[width=\textwidth]{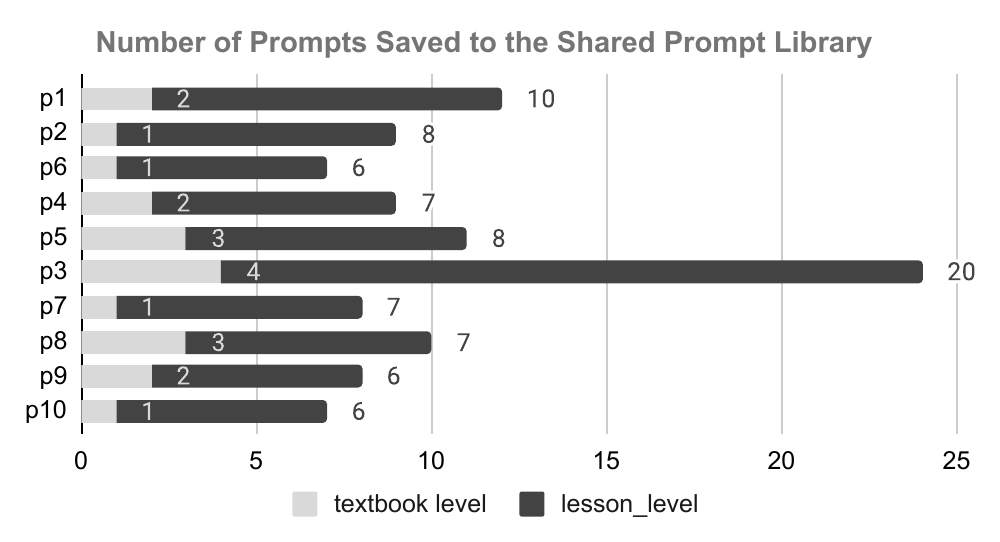}
    \caption{The number of times participants \textbf{saved} prompts.}
    \label{fig:commits} 
  \end{minipage}
\end{figure}

The PromptHive system logging engine captured data on user interactions whenever subject matter experts executed a prompt or saved it to the shared library. Figure \ref{fig:commits} and \ref{fig:executions} shows the distribution of executions and saves across the ten subject matter experts, indicating that on average, they made 30 executions ($SD=12.57$) and 10 commits each ($SD=5.02$) to author the 10 textbook-level prompts and 59 lesson-level prompts covering hints for the entirety of a textbook. The logging engine also links the prompts captured into a tree structure that can be exported as a JSON file. See Figure \ref{fig:logging} for a snapshot of the rich interactions captured during the 1.5 hour sessions. 

From this log, we found that participants did between 1 (P2) and 17 (P3 and P8) iterations based on executions to arrive at their final textbook-level prompt, with most taking between 3 to 6 iterations. Below is an example of the textbook-level chain from P5, with the removals and additions across each iteration highlighted in red and blue, respectively:

\begin{figure*}
  \centering
  \includegraphics[width=\linewidth]{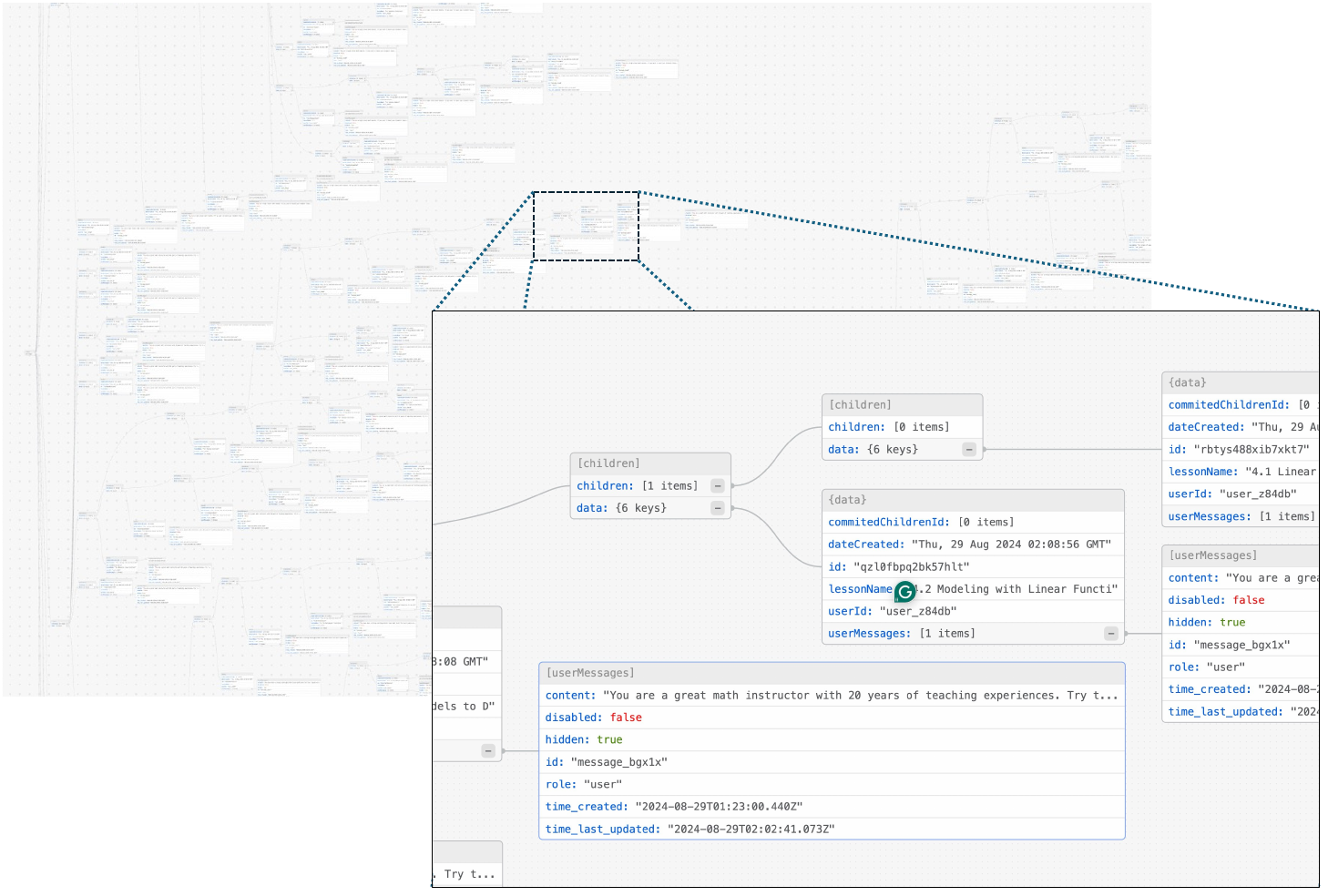}
  \caption{A snapshot of the JSON tree structure captured by PromptHive’s logging engine, illustrating how researchers can retrace the iterative process of domain experts when refining prompts. Each {data} node in this example represents an execution of a prompt variation within the scratchpad. A similar logging mechanism tracks how prompts are committed to the shared prompt library.}
  \label{fig:logging} 
\end{figure*}

\begin{itemize}
    \item \textbf{Iteration 1:} You have 20 years of experience in teaching high school math and middle school math and specialized in helping special education students understand the math contents. Currently, you are working with a group of ELD students who have ADHD and severe learning disability to understand math. Make sure the hints that you give are concise which means less than 10 words for each step. They are also easy to understand, encouraging, and interesting for students to follow. Those students are also historically known to be marginalized.
    
    \item \textbf{Iteration 2:} You have 20 years of experience in teaching high school math and middle school math and specialized in helping special education students understand the math contents. Currently, you are working with a group of ELD students who have ADHD and severe learning disability to understand math. Make sure the hints are enthusiastic, easy to understand, encouraging, and interesting for students to follow. \textcolor{red}{Make sure the hints that you give are concise which means less than 10 words for each step.} \textcolor{red}{They are also historically known to be marginalized.}
    
    \item \textbf{Iteration 3:} You have 20 years of experience in teaching high school math and middle school math and specialized in helping special education students understand the math contents. Currently, you are working with a group of ELD students who have ADHD and severe learning disability to understand math. Make sure the hints are enthusiastic, easy to understand, encouraging, and interesting for students to follow. \textcolor{red}{No change from Iteration 2.}
    
    \item \textbf{Iteration 4:} You have 20 years of experience in teaching high school math and middle school math and specialized in helping special education students understand the math contents. Currently, you are working with a group of ELD students who have ADHD and severe learning disability to understand math. Make sure the hints are enthusiastic, easy to understand, encouraging, and interesting for students to follow. \textcolor{red}{No change from Iteration 3.}
    
    \item \textbf{Iteration 5:} You have 20 years of experience in teaching high school math and middle school math and specialized in helping special education students understand the math contents. Currently, you are working with a group of \textcolor{blue}{special education} students. Make sure the hints are enthusiastic, easy to understand, encouraging, and interesting for students to follow. \textcolor{red}{ELD students who have ADHD and severe learning disability to understand math.}
    
    \item \textbf{Final Prompt:} You have 20 years of experience in teaching high school math and middle school math and specialized in helping special education students understand the math contents. Currently, you are working with a group of special education students. \textcolor{blue}{You need to add some emojis to each hint to make it interesting for students to follow.} Make sure the hints are enthusiastic, easy to understand, encouraging, and interesting for students to follow.
\end{itemize}

Here, we see how P5 starts off by experimenting with writing hints for students with learning disabilities, even though the focus of the textbook was for college-level math, likely drawing from their experience in teaching  diverse students as a 7th grade math teacher. In iteration two, they remove the last two sentences on making the hints concise and being mindful of the marginalization of such students. Given how LLMs can generate different outputs for executions of the same prompt, we see there was no change to the content from Iteration 3 to 4. In 5, they mention special education twice, likely to emphasize who they want these hints to target. Finally, they add a note about including emojis to make the hints more engaging. The JSON log captures similar chains for both textbook-level and lesson-level prompts across all participants, which we cannot detail here due to space considerations but will attach as supplementary material as it offers empirical insights on how subject matter experts iterate prompts. See Appendix \ref{appendix:prompts} for a list of finalized textbook level prompts from all participants.

Looking at how participants' textbook-level prompts influenced each others' lesson-level prompts, as shown in Figure \ref{fig:collaborate}, we find that upvotes did not correlate with influence, as the most influential post from P8, which influenced 4 other participants, had 0 upvotes. Apart from P3's prompt, which did not influence any lesson-level prompt, most participants influenced at least one group of lesson-level prompts (often their own), and sometimes those of several others, in alignment with user comments on collaboration being helpful in prompt-authoring. Of the 59 lesson-level prompts committed to PromptHive, only 8 were verbatim textbook-level clones, i.e., they didn't need tailoring. The domain experts chose to tailor the remaining 51.

\section{Study 2: Learning Gain Study with College-Level Math Learners}
In this study, we explore the impact of PromptHive hints on learning gains compared to a human-only authored hints control condition.

\subsection{Research Questions}
    \begin{itemize}
        \item RQ5: Do PromptHive-generated hints lead to learning gains, and how do those gains compare to those from human-only hints?
    \end{itemize}

\subsection{Participants}
Through Prolific, we recruited a total of 358 current undergraduate college students, with 266 unique participant submissions for the human-only authored hints control condition and 263 for the PromptHive experiment condition. Since participants completed a sequence of 3 lessons, we excluded any lesson submissions where participants did not fully complete all parts of the lesson sequence (i.e. 3-question pre-test, 5-question hint condition, 3-question post-test). After this exclusion, we resulted in 225 unique participants, with a total of 549 completed lesson submissions (268 for human-only and 281 for PromptHive).

\subsection{Tasks}
Participants completed a 3-question pre-test to assess their initial understanding of the lesson’s topic. They were then presented with 5 additional questions, receiving correctness feedback on their answers along with a tutoring pathway based on the assigned condition (either a human-curated hint pathway or the PromptHive-generated pathway). Finally, participants completed a post-test with the same 3 questions as the pre-test to assess learning gains. This sequence was repeated for 3 distinct lesson/condition pairings. At the end of the experiment, participants were shown a survey code and asked to enter it into their Prolific portal. 

\subsection{Materials}
For the human-only control hints, we use the hints generated by the OATutor project, contained within the OATutor system. These hints were multi-level hint pathways, curated by subject matter experts with prior tutoring experience. There was no restriction on the number of hints and scaffolds to incorporate in the hint pathway. 

The prompt utilized for PromptHive to generate hint pathways for all questions in a single response incorporated the formatting guidelines provided to OATutor's content curation team. These guidelines specified detailed instructions for rendering problems correctly in the OATutor system, with mathematical expressions following the required syntax for proper display and functionality. Each response generated by PromptHive was a JSON object with 80 keys, one for each question. The value for each key was a string that, when parsed, contained a multi-hint pathway specific to that question. Building on the methodology of a study that employs self-consistency \cite{wang2022self} to reduce hallucinations for worked solutions in mathematics, achieving near 0\% hallucination in College Algebra \cite{pardos2024chatgpt}, we used a similar approach for hallucination mitigation. Since each response string by PromptHive contained a multi-hint pathway rather than a worked solution, we used a similar approach to hallucination mitigation, but one which involved vectorizing all the responses to a single question and finding the most representative response (closest to the centroid) \cite{li2024more}. In essence, we prompted PromptHive to generate this JSON object 30 times, resulting in 30 versions of the multi-hint pathways for each of the 80 questions. For each set of 30 responses for a single question, we first vectorized the responses using SentenceTransformer and found the centroid vector. Then, we used cosine similarity to find the response closest to the centroid vector, which represents the most representative response out of our 30. This response was then used in our study and put into the OATutor system as being the multi-hint pathway for the PromptHive condition. 

In order to ensure compatibility with OATutor, we checked that each of the 80 questions was able to be rendered correctly in the system, complying with its formatting guidelines. This check resulted in having to make slight modifications to the PromptHive generated responses, such as changing questions with multiple correct answers to “multiple choice” and “string” (exact-match) answer types. These modifications were made by the second and third authors of the paper, taking them 3 hours and 15 minutes and 6 hours and 19 minutes to complete, respectively. Sample hint pathways for both the PromptHive hints and human-only control hints are shown in Appendix \ref{sec:sample_hint_pathways}.

\subsection{Procedure}
We used the Qualtrics platform to facilitate random assignment of learners to either the human-only authored hints control or the PromptHive experiment condition. Prolific, a crowdsourcing platform, was used to recruit current undergraduate college students in the United States to complete the study. The participants were compensated with \$20 USD for study completion. All participants first viewed an instructions screen which explained the task, after which they were randomly assigned to one of twenty lesson/condition configurations (10 lessons, 2 conditions).

The OATutor system’s log data was used to track user actions, including inputting answers, opening hints, conditions assigned, timestamps, and other interactions.

\subsection{Analysis}
We assessed the normality of average pre-test scores, post-test scores, learning gains, and time-on-task using the Shapiro-Wilk Test of Normality. Since we rejected the null hypothesis (indicating that the data was not normally distributed), we used the Kruskal-Wallis test to analyze pre-test scores and time-on-task to assess evenness at pre-test and detect statistically significant differences in time spent per lesson. Subsequently, we utilized the Mann-Whitney U test for pairwise comparisons. If we had failed to reject the null hypothesis of the Shapiro-Wilk Test of Normality (indicating normality), we would conduct the same analysis using ANOVA. 

To examine pre- to post-test learning gains for the lesson and condition pairings, we used the Wilcoxon signed-rank test for paired samples since the data was not normally distributed. Otherwise, we would have utilized a paired t-test for the same comparisons. Finally, for understanding how the hint conditions compare to one another, we utilized a mixed linear model (with ranked data due to non-normality), incorporating fixed effects of the condition and random effects for the participant. Specifically, the following mixed-effects model was used in the analysis:

\[
\text{Learning Gain}_{i} = \beta_0 + \beta_1 \cdot \text{condition}_i + u_j
\]

where \(\beta_0\) is the intercept, \(\beta_1\) is the fixed effect of the condition, \(u_j\) is the random effect for participant \(j\), and \(\epsilon_i\) is the error term.

\subsection{Results}
The Shapiro-Wilk Test indicated non-normality for average pre-test scores, post-test scores, learning gains, and time-on-task. Therefore, we further utilized Kruskal-Wallis and failed to reject the null for all lessons and thus found both hinting conditions to be even at pre-test for each lesson, allowing for sound comparisons. Average pre-test scores were 51.9\% for the PromptHive hints condition and 45.9\% for the human-only control. When comparing time-on-task between the two hint types, the Kruskal-Wallis test revealed statistically significant differences overall (p = 0.031). Through the Mann-Whitney U test, we found that only lesson 8.4 showed statistically significant differences between time-on-task of the PromptHive hints and human-only authored hints (p = 0.030). 

\begin{figure*}
  \centering
  \includegraphics[width=\linewidth]{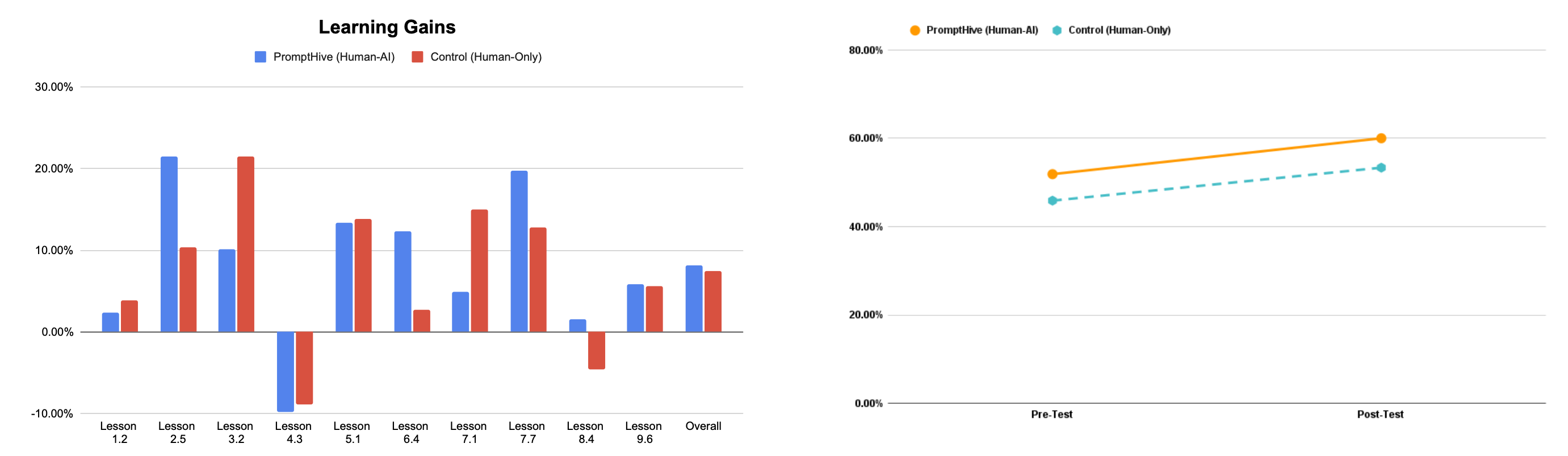}
  \caption{Learning gains by lesson and overall pre- to post-test scores.}
  \label{fig:learning_gains}
\end{figure*}

Next, we utilized the Wilcoxon signed-rank test for paired samples and found both the PromptHive (p < 0.001) and the control (p < 0.001) conditions to exhibit statistically significant differences, indicating that both conditions facilitate learning. Table \ref{learning_gains} shows these results granularly at the lesson level, with 4 lessons showing statistically significant learning gain with PromptHive compared to 2 lessons for the human-only authored hints. Average learning gain in the control condition was 7.47\% and for PromptHive was 8.13\%, with 3 lesson/condition pairings exhibiting negative learning gains (4.3 PromptHive, 4.3 control, and 8.4 control). Figure \ref{fig:learning_gains} shows these learning gains. Due to non-normality, we utilized the mixed linear model described previously with ranked data. We found no statistically significant differences between the learning gains of the hint conditions (p = 0.688).

\begin{table}[ht]
\centering
\scriptsize
\begin{tabular}{llcccccc}
\toprule
Lesson & Condition & N & Avg. Time (s) & Learning Gain (\%) & Avg. Pre-test (\%) & Avg. Post-test (\%) & p-value \\
\midrule
\multirow{2}{*}{1.2} & PromptHive & 29 & 480.0  & 2.31\%  & 74.71\% & 77.02\% & 0.315 \\
                     & Control    & 26 & 585.5  & 3.85\%  & 65.38\% & 69.23\% & 0.224 \\
\multirow{2}{*}{2.5} & PromptHive & 31 & 663.0  & 21.50\% & 47.31\% & 68.81\% & \textbf{0.027} \\
                     & Control    & 29 & 533.0  & 10.34\% & 37.93\% & 48.28\% & 0.235 \\
\multirow{2}{*}{3.2} & PromptHive & 33 & 449.0  & 10.11\% & 50.51\% & 60.62\% & \textbf{0.024} \\
                     & Control    & 28 & 312.0  & 21.44\% & 41.66\% & 63.11\% & \textbf{0.001} \\
\multirow{2}{*}{4.3} & PromptHive & 34 & 579.0  & -9.80\% & 80.39\% & 70.59\% & \textbf{0.026} \\
                     & Control    & 30 & 459.5  & -8.88\% & 74.45\% & 65.56\% & 0.225 \\
\multirow{2}{*}{5.1} & PromptHive & 25 & 1380.0 & 13.34\% & 42.66\% & 56.00\% & 0.147 \\
                     & Control    & 29 & 748.0  & 13.80\% & 34.47\% & 48.27\% & 0.068 \\
\multirow{2}{*}{6.4} & PromptHive & 31 & 667.0  & 12.37\% & 54.30\% & 66.67\% & 0.159 \\
                     & Control    & 25 & 444.0  & 2.67\%  & 56.00\% & 58.67\% & 0.725 \\
\multirow{2}{*}{7.1} & PromptHive & 27 & 755.0  & 4.95\%  & 66.66\% & 71.61\% & 0.302 \\
                     & Control    & 29 & 1025.0 & 14.93\% & 63.23\% & 78.16\% & \textbf{0.025} \\
\multirow{2}{*}{7.7} & PromptHive & 27 & 1325.0 & 19.76\% & 27.15\% & 46.91\% & \textbf{0.015} \\
                     & Control    & 26 & 1225.0 & 12.83\% & 30.76\% & 43.59\% & 0.112 \\
\multirow{2}{*}{8.4} & PromptHive & 21 & 880.0  & 1.60\%  & 52.38\% & 53.98\% & 0.753 \\
                     & Control    & 22 & 624.0  & -4.55\% & 43.94\% & 39.39\% & 0.418 \\
\multirow{2}{*}{9.6} & PromptHive & 23 & 1566.0 & 5.79\%  & 7.24\%  & 13.03\% & 0.102 \\
                     & Control    & 24 & 1430.5 & 5.55\%  & 4.16\%  & 9.71\%  & 0.102 \\
\bottomrule
\end{tabular}
\caption{Learning gain results.}
\label{learning_gains}
\end{table}
\section{Discussion}

In this paper, we introduced PromptHive, an open-source collaborative prompt authoring interface that integrates seamlessly with the content authoring workflow of an open adaptive tutoring system, OATutor \cite{pardos2023oatutor}. Through a review of frameworks for Human-AI collaboration and prompt-authoring interfaces for LLMs, we developed four grounding design principles -- Integration, Simplicity, Trust, and Collaboration -- to augment the expert workflow of subject matter experts in OATutor. Through iterative brainstorming sessions with the content team lead over 6 months, we developed a high-fidelity prototype that implements a four-stage prompt-authoring workflow: Load, Author, Share, Iterate.

Rather than displacing content authors, PromptHive involved them deeply in the prompt authoring process for collaboratively creating AI-generated problem-help at a greater scale, while retaining control over the substance and style of the hints. This aligns with Shneiderman's optimistic view of human-centered AI automation \cite{shneiderman2022}, where AI systems can increase automation while preserving human control. Our evaluation, which involved 10 subject-matter experts, and over 200 undergraduate learners, reflected this goal. The experts agreed on the system’s high usability and trustworthiness, expressing enthusiasm about integrating AI in ways that complemented their existing human authoring process. As P8 noted, the experts did not feel replaced by the AI; instead,  they highlighted the advantages of incorporating PromptHive into their content creation workflow.

PromptHive’s logging engine captured empirical data on how experts iterated on prompts, providing granular insights that would have been difficult to obtain through interviews alone. For instance, we observed how P5 experimented with prompts aimed at younger students with special needs before finalizing a prompt for a college-level audience. Their focus on tone, including the use of encouraging language and emojis, resonated with many other participants, who remarked on these features during post-interviews. Although upvotes did not correlate with influence, future work could explore other motivational factors in prompt selection, as evidenced by the fact that four other experts adopted P8’s prompt, despite it having 0 upvotes. 

To further assess PromptHive’s efficacy, we conducted a second study comparing prompts generated by experts using PromptHive with those written by past experts without AI support. Both sets of prompts produced comparable learning gains, which is notable considering that PromptHive reduced the time required by many folds and the perceived cognitive load required by more than half. This increase in efficiency did not come at the expense of experts sense of autonomy and control. The speedup could have been more dramatic by involving the editors in system-level prompt design, which would reduce the time spent correcting rendered output and addressing periodic syntax issues in some of the generated hint pathway outputs.


The four-part workflow we designed and evaluated in PromptHive -- \textit{loading} structured content (initially from an existing human-only expert workflow), \textit{authoring} prompts that generate AI content integrated with the structured content, giving experts an opportunity to view the output in context, \textit{sharing} the most effective formulations with other subject matter experts so they can borrow ideas from them and refine them further through continuous cycles of \textit{iteration} -- could be applied to other contexts such as educational curricula. Given the outlines of individual lessons, the four-part workflow could potentially serve as a method of rapidly generating class curriculum, lending itself to the pre-existing well-defined nature of the content's structure. 

This PromptHive workflow could conceivably be abstracted to any scenario involving text production micro-tasks that require consideration of contextual information and structured output. In these scenarios, there is value in migrating human labour away from this type of tedious work to instead authoring and maintaining policies that effectively generalize across related contexts. In such scenarios, this workflow promotes the worker from a doer role to a managerial role, embodying the producer-to-curator shift.
This would not include long form writing tasks, such as authoring a novel, but could include writing up expense reports or responding to customer support emails \cite{ashktorab2021}, where representatives become prompt engineers focused on authoring general prompts (i.e., "textbook-level prompts") to respond to customer queries, as well as tailoring prompts for dealing with particular sub-categories of support issues (i.e., "lesson-level" prompts). Much like our scenario where we had a collection of manually authored hints and participants with experience writing them, such customer service reps could use existing emails that they have received and responded to in order to evaluate how well the PromptHive generated email responses compare in terms of likely customer satisfactions or problem resolution scores, akin to our learning gain scores.

For such workflows, thoughtful AI integration is crucial. AI should adapt to the needs and nuances of the expert workflow, rather than forcing experts to adjust to the limitations of the LLM. This contrasts with many AI integrations, where the LLM operates as a side tool, typically in the form of a chat interface, processing unstructured input and producing unstructured output. By incorporating recent advancements in LLMs, such as vision capabilities in GPT-4o to read images and new APIs that ensure adherence to specific data schemas, we designed PromptHive to more closely align with the structured nature of the authoring workflow.



\section{Conclusion}





We designed PromptHive in collaboration with ten mathematics tutoring subject matter experts to allow them to author prompts, immediately see LLM outputs across a number of different problem contexts, and then save these outputs within the native publishing workflow of an adaptive tutoring system project. Our study of the subject matter experts found that PromptHive cut NASA-TLX subjective cognitive load in half, from 55.17 to 26.73, and reduced the total time to author hints by factor of 30. This reduction in workload and increase in efficiency was accomplished without a decrease in content quality, as measured by our randomized controlled study of 225 learners engaging with 10 lessons and showing significant learning gains from PromptHive produced hints (8.13\%, p < 0.001) that were not statistically significantly different from the learning produced by the more laborious and time consuming manual hint authoring process (7.47\%) that did not utilize PromptHive (p = 0.688).



Participants remarked on how much they valued the collaborative aspect of the system, learning from each other's prompt ideas while retaining editorial control of the produced content. This interest in prompt improvement and retaining their expert judiciousness was further made evident when participants were given the option to use one of the existing shared textbook-level prompts verbatim, but in 86\% of the lessons instead chose to further tailor a prompt to the lesson. These positive user experiences were reflected in an "Excellent" overall average system usability score of 89/100.  

Around 40\% of the total manual time spent producing hints was spent by editors on correcting relatively minor syntactic issues in the output of PromptHive. Future design revisions could involve editors focused on rendering and syntax, rather than pedagogy, in system-level prompt iteration to reduce the occurrence of these issues.

PromptHive demonstrates a way forward for human-controlled AI content production that migrates workers from strictly producers to curators and shows how, with effective design, LLMs can improve task efficiency as well as domain expert satisfaction.

\begin{acks}
We thank the Vice Provost of Undergraduate Education's Micro Grant Program for providing financial support for this work. We also extend our gratitude to Joe Fang and Sarva Sanjay from the Department of Computer Science at the University of Toronto for their contributions, with Joe assisting in deploying the prototype and Sarva implementing the back-end logging engine.
\end{acks}

\bibliographystyle{ACM-Reference-Format}
\bibliography{references}

\appendix
\clearpage
\section{Trust Scale}
\label{appendix:trust}

\begin{table}[htbp]
\centering
\scriptsize
\begin{tabular}{|l|}
\hline
1. I am confident in PromptHive. I feel that it works well. \\ \hline
2. The outputs of PromptHive are very predictable. \\ \hline
3. PromptHive is very reliable. I can count on it to be correct all the time. \\ \hline
4. I feel safe that when I rely on PromptHive I will get the right answers. \\ \hline
5. PromptHive is efficient in that it works very quickly. \\ \hline
6. I am wary of PromptHive (adopted from the Jian, et al. Scale and the Wang, et al. Scale). \\ \hline
7. PromptHive can perform the task better than a novice human user (adopted from the Schaefer Scale). \\ \hline
8. I like using PromptHive for decision making (adopted from the Madsen-Gregor Scale). \\ \hline
\end{tabular}
\caption{XAI context trust scale adapted from \cite{hoffman2023measures}.}
\label{tab:promptHiveStatements}
\end{table}

\clearpage

\section{Sample Hint Pathways}
\label{sec:sample_hint_pathways}

\begin{figure}[ht]
    \centering
    \fbox{\includegraphics[width=1\textwidth]{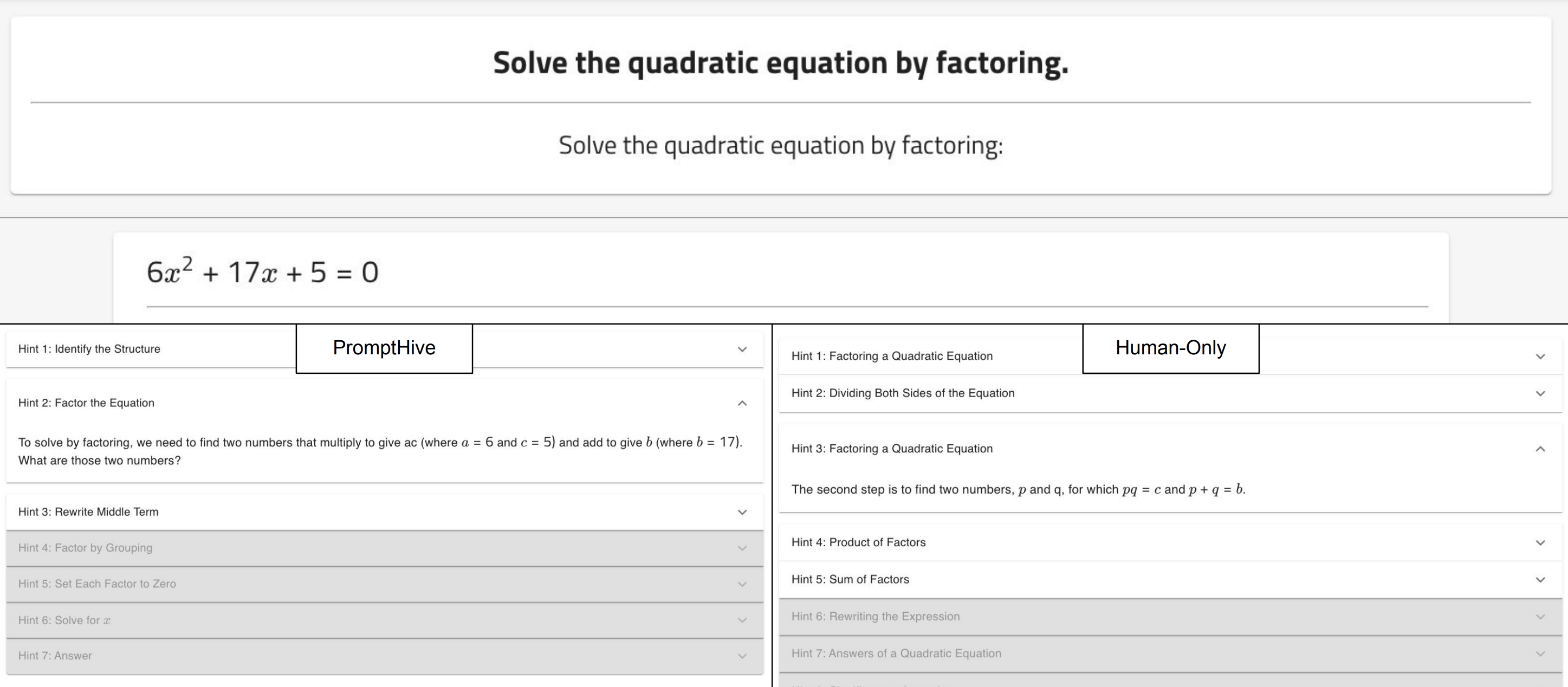}}
    \caption{Sample hint pathways for PromptHive and human-only hints in lesson 2.5.}
    \label{fig:lesson2.5}
\end{figure}

\begin{figure}[ht]
    \centering
    \fbox{\includegraphics[width=1\textwidth]{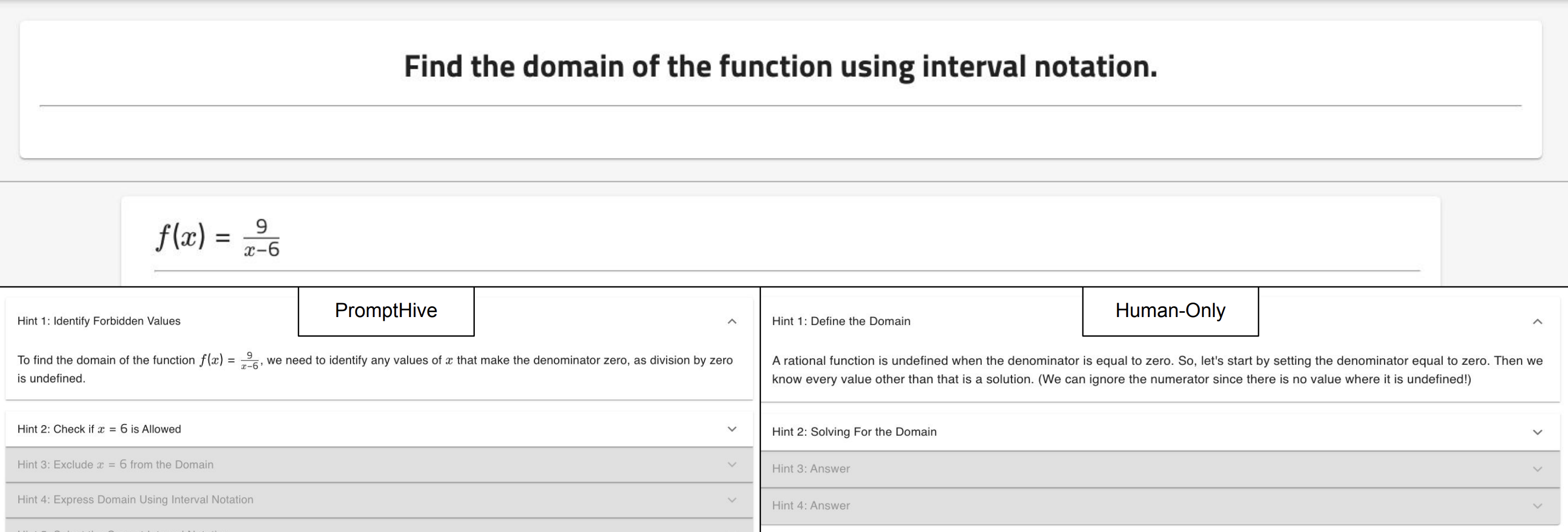}}
    \caption{Sample hint pathways for PromptHive and human-only hints in lesson 3.2.}
    \label{fig:lesson3.2}
\end{figure}

\begin{figure}[ht]
    \centering
    \fbox{\includegraphics[width=1\textwidth]{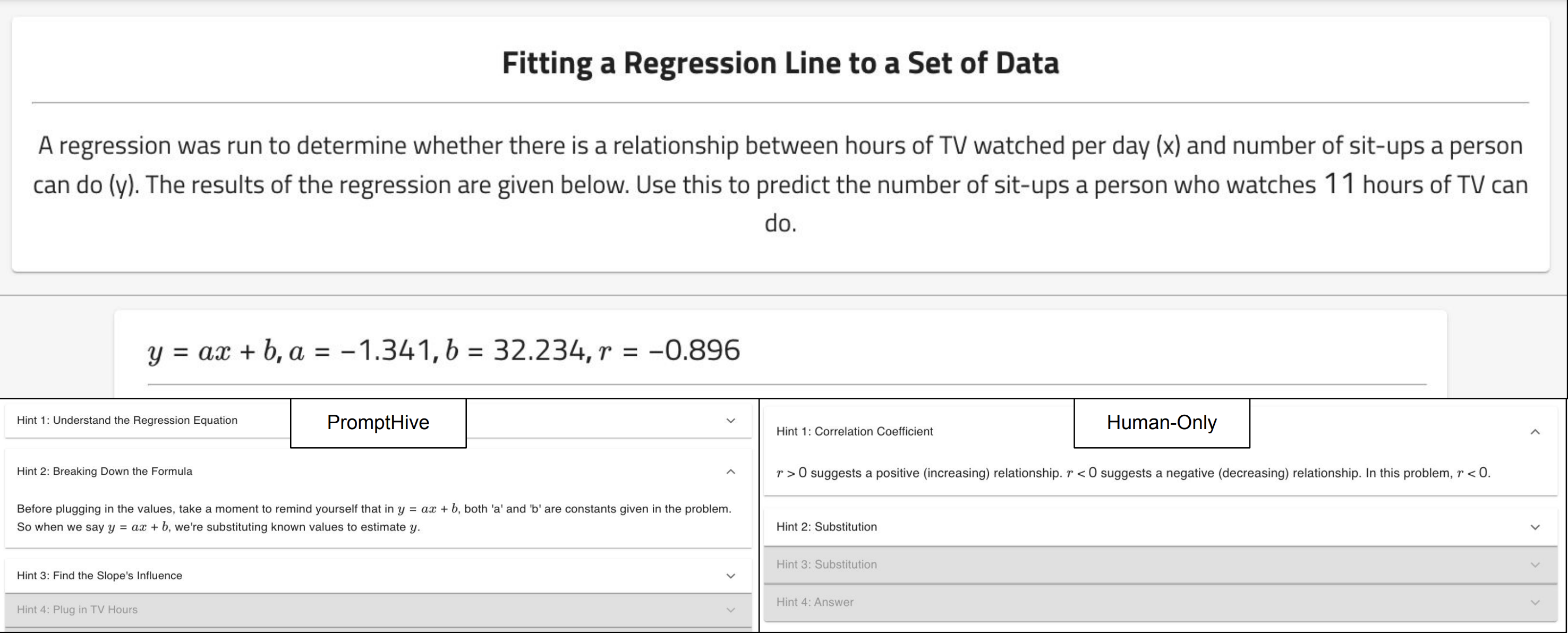}}
    \caption{Sample hint pathways for PromptHive and human-only hints in lesson 4.3.}
    \label{fig:lesson4.3}
\end{figure}

\begin{figure}[ht]
    \centering
    \fbox{\includegraphics[width=1\textwidth]{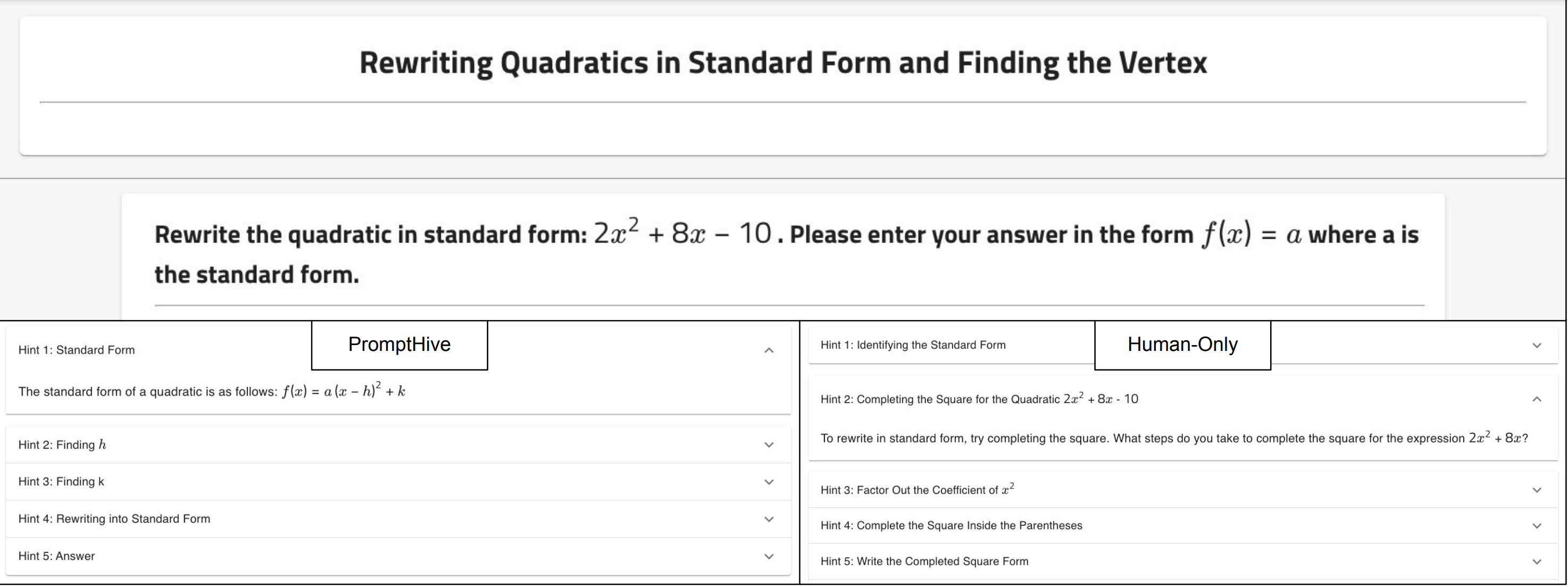}}
    \caption{Sample hint pathways for PromptHive and human-only hints in lesson 5.1.}
    \label{fig:lesson5.1}
\end{figure}

\clearpage

\section{Finalized Textbook-level Prompts in Study 1}
\label{appendix:prompts}

\begin{longtable}{|c|p{12cm}|c|}
\hline
\textbf{Participant ID} & \textbf{Prompt User Message} \\ \hline
p1 & \small{You are a high school math teacher. If you want to teach your students these questions, how would you break down each problem with meaningful hints to help them effectively learn the material? Try not to have repeated hints. Try to have a positive tone!}\\ \hline
p2 & \small{You are a college math professor tutoring a new college student. Your plan is to create a set of hints to help the student understand the problem. Include at least 2 hints and 1 scaffold for each problem. Begin the series of hints with general hints and slowly create hints that are more specific to the problem. Avoid asking questions at the end of hints. Make sure to explain concepts and properties.}\\ \hline
p3 & \small{You are a math tutor instructing college algebra helping a student with understanding algebra. Create hints to help the student solve the following problem. Remember that scaffolds are smaller parts of the main question that the student will answer, and hints are statements that help guide the student to think and answer a scaffold or the main question. Make the later hints more simple as the question and its hints goes on.}\\ \hline
p4 & \small{Ignore previous instructions. You are a math teacher who is trying to explain some problems to students. When looking at each question, give the students some hints that would allow them to solve their problem. DO NOT reveal the answer. DO NOT repeat the question in your hints. DO NOT repeat information from hint to hint. In your hints, try to prioritize explaining the theory behind the topic they are learning before diving into solving into the question. Your hints can also ask the students questions. Try to be positive and helpful. In your scaffolds, try to answer smaller parts of the question. If the question is too simplistic, it's not necessary to involve scaffolds. Ensure that these scaffolds do not give away the answer to the question entirely. For scaffolds, ensure the answer type is numeric or multiple choice; avoid long input answers or string answers.}\\ \hline
p5 & \small{You have 20 years of experience in teaching high school math and middle school math and specialized in helping special education students understand the math contents. Currently, you are working with a group of special education students. You need to add some emojis to each hints to make it interesting for students to follow. Make sure the hints are enthusiastic, easy to understand, encouraging, and interesting for students to follow.}\\ \hline
p6 & \small{You are a great math instructor with 20 years of teaching experiences. Try to give out hints or scaffolds to students, help them understanding the underlying concepts without giving out the direct answers. Also try to explain the mathematical terms to students as 7 years old kids, make it as easy as possible for them to understand. Walk them over the entire thought process, so they can solve similar problems themselves in the future.}\\ \hline
p7 & \small{You are a math teacher of 10 years with a deep understanding of many different mathematical concepts, but is renown for explaining how to solve problems in layman terms so that students past the 8th grade are able to understand easily. Create hints for the problems above and work through the problems, but don't give out any direct answers until the final hint. Try using leading questions instead of directly telling them what the answers are for each step.}\\ \hline
p8 & \small{You are a college-level math instructor, who is descriptive, but concise. Please provide hints and scaffolds for these questions, but you should not in any way give away the solution to students. The difference between a hint and a scaffold is a hint is a conceptual guide to approaching the question, while a scaffold ends with a question mark ""?"" and asks the student to solve for a technical part of the question (for instance, what the simplified fraction looks like). In your hints and scaffolds, do not use any fancy jargon; please maintain a friendly, but professional demeanor. Please provide 1-3 hints and 2-5 scaffolds per question, where the first hint is a relatively broad, conceptual hint, and as you progress through each question, there are less hints and more scaffolds, making them more specific to each step of the process.
Your goal in each hint and scaffold is to make sure the student understands a little more than they did before reading (and working through) that hint. Can you provide an explanation at each step about why the student needs to perform that step?}\\ \hline
p9 & \small{You are a patient, friendly math tutor with 20 years of experience who wants to help students solve problems by giving them hints. Generate hints that will help a student solve the problems and also understand a general logic to the concept. Title case all the hints. 

For hints, avoid questions with ""?"" and instead share what is a good way to proceed with the question. For hints, make the titles start with some actionable non-form word, such as ""Identify XX"" 

For scaffolds, include specific numbers that are only applicable to the question so that students can start getting more concrete progress. For scaffolds, make the answer type to be numeric or multiple choice. Avoid long equation input answers as they are hard to type.}\\ \hline
p10 & \small{You are a tutor for a college-level math class. Make sure to be friendly and welcoming. Your goal is to create a set of hints for questions that scaffold and come one after the other. Your hints should start off simple and should aim to guide students into the right direction as opposed to giving them the answer straight away. Start off by asking questions as hints that will help students understand what the problem is asking them. As you give more hints give students some practice problems that will help them understand what you are trying to teach them. At the end of the problem make sure to give the answer. Remember that your main goal is to teach the students and help them understand what they are being asked to do and how to do it.}\\ \hline
\caption{Collection of finalized textbook-level prompts from subject matter experts in Study 1.}
\end{longtable}

\end{document}